\documentclass[aps,pra,10pt,notitlepage,nofootinbib,tightenlines,floatfix,
twocolumn,superscriptaddress]{revtex4-2}

\usepackage[margin=1in]{geometry}

\usepackage[colorlinks=true, allcolors=blue]{hyperref}
\hypersetup{
    colorlinks=true,       
    linkcolor=red,          
    citecolor=magenta,        
    filecolor=magenta,      
    urlcolor=cyan,           
    runcolor=cyan
}
\usepackage{bm,bbm,amssymb,amsmath,amsfonts,amsthm,mathrsfs,MnSymbol,times}
\usepackage{natbib}
\usepackage{graphicx}
\usepackage{amsmath}
\usepackage{mathtools}
\usepackage{multirow}
\usepackage{color}
\usepackage{bbold}
\usepackage[utf8]{inputenc}
\usepackage{empheq}
\usepackage{soul}
\usepackage[ampersand]{easylist}
\usepackage[normalem]{ulem}
\usepackage{braket}
\usepackage{enumerate}
\usepackage[shortlabels]{enumitem}
\usepackage{blkarray}
\usepackage{tabularx}
\usepackage{printlen}
\usepackage{siunitx}

\newcommand{\bel}{\begin{easylist}[itemize]}
\newcommand{\eel}{\end{easylist}}

\newcommand{\kyiv}[0]{\texttt{ibm\_kyiv}}
\newcommand{\sherbrooke}[0]{\texttt{ibm\_sherbrooke}}
\newcommand{\marrakesh}[0]{\texttt{ibm\_marrakesh}}
\newcommand{\osaka}[0]{\texttt{ibm\_osaka}}
\newcommand{\kolkata}[0]{\texttt{ibmq\_kolkata}}
\newcommand{\cairo}[0]{\texttt{ibm\_cairo}}
\newcommand{\bogota}[0]{\texttt{ibm\_bogota}}
\newcommand{\jakarta}[0]{\texttt{ibm\_jakarta}}

\graphicspath{ {figures/} }

\begin{document}

\title{Crosstalk-Robust Dynamical Decoupling for Bipartite-Topology Quantum Processors}

\author{Ethan Hickman}
\affiliation{Joint Center for Quantum Information and Computer Science (QuICS), Department of Computer Science\\ University of Maryland, College Park, College Park, Maryland 20742, USA}

\author{Xiaodi Wu}
\affiliation{Joint Center for Quantum Information and Computer Science (QuICS), Department of Computer Science\\ University of Maryland, College Park, College Park, Maryland 20742, USA}

\author{Gregory Quiroz}
\affiliation{Johns Hopkins University Applied Physics Laboratory,
Laurel, Maryland 20723, USA}
\affiliation{William H. Miller III Department of Physics \& Astronomy,
Johns Hopkins University, Baltimore, Maryland 21218, USA}

\begin{abstract}
We introduce a protocol that modifies dynamical decoupling (DD) sequences to be robust to static $ZZ$ crosstalk when implemented with bounded control on two-colorable qubit topologies. The protocol, which relies on modifications to the pulse timing, can be applied to any sequence with equidistant $\pi$-pulses. We motivate the method theoretically via suppression conditions identified through time-dependent perturbation theory. Theoretical findings are supported by demonstrations of widely studied sequences on several superconducting qubit devices offered by the IBM Quantum Platform. Using up to 20 qubits on fixed-coupler devices, we observe at least a $3\times$ improvement in the fidelity decay rate via our approach when compared to non-robust DD variants. In addition, we leverage our approach to assess the impact of $ZZ$ errors on tunable-coupler devices. We find that $ZZ$-robust sequences perform nearly equivalent to non-robust DD, affirming the reduced impact of such errors in a tunable-coupler architecture. Nevertheless, our demonstrations indicate that fixed-coupler devices, when subject to DD-protection, can outperform tunable-coupler devices. Our method broadens the scope of practical DD protocols: with modest overhead and a reasonable constraint on the qubit topology, the method attains significant performance improvements on modern quantum computing devices.
\end{abstract}

\maketitle

\section{Introduction}
\label{sec:intro}
Dynamical decoupling (DD) is a powerful tool for achieving error suppression in quantum computing and sensing applications. While DD has a long history, dating back to nuclear magnetic resonance~\cite{hahn1950spinechoes, carr1954effectsdiffusion, meiboom1958modifiedspinecho, maudsley1986modifiedcarrpurcellmeiboomgill,haeberlen2012high}, recent demonstrations have further illustrated its utility in modern quantum computing devices. For example, in superconducting qubit systems, DD has proven to be an effective technique for preserving single- and multi-qubit quantum states~\cite{pokharel2018demonstrationfidelity, tripathi2022suppressioncrosstalk, ezzell2023dynamicaldecoupling, zhou2023crctrl, rahman2024learninghowa, tong2025empiricallearning}. In addition, boosts in the accuracy of quantum algorithms~\cite{ravi2021vaqemvariational, niu2022effectsdynamical, kim2023scalableerror, pokharelDemonstrationAlgorithmicQuantum2022, pokharel2024betterthanclassicalgrover, ji2024synergisticdynamical, singkanipa2024demonstrationalgorithmic} and passive~\cite{quiroz2024dfs,Mena-Lopez:2023aa,han2024protecting} and active quantum error correction~\cite{chen2021qec, krinner2022realizing,ai2023suppressing,postler2023demonstration,bluvstein2023logical} have been realized using DD. The utility of DD is not limited to near-term systems, and is anticipated to continually demonstrate value in the fault-tolerant quantum computing regime as well~\cite{ng2011dd}.

DD accomplishes error suppression by using dynamical manipulation of a quantum system via strong and fast pulses to selectively average out unwanted interactions between a quantum system and its environment. While DD is not the only tool available to improve the reliability of quantum devices, it does have some advantages over other error management techniques. For example, quantum error mitigation (QEM)~\cite{cai2023quantumerror} refers to a set of techniques that apply statistical arguments to reduce the impact of noise on estimates of observables. In contrast to DD, QEM methods often rely on sampling and classical post-processing that can scale unfavorably with increasing system size~\cite{takagi2022fundamental, takagi2023qem}. Being an open-loop technique, DD has advantages over quantum error correcting codes (QECCs)~\cite{shor1995schemereducing, steane1996errorcorrecting, gottesman1997stabilizercodes, bravyi1998quantumcodes, terhal2015quantumerror} as well. QECCs relies on active feedback to detect and rectify errors. However, this process demands high-fidelity operations, long coherence times, and fast classical feedback to reach anticipated fault-tolerant operating conditions~\cite{gottesman2016quantumfaulttolerancesmall,shor1996fault}.

There exists a vast ecosystem of DD protocols meant to support a wide variety of control and noise scenarios. Idealizations of control, in which the pulses are error-free and instantaneous (i.e., infinite in amplitude), are commonly used to identify and derive protocols that achieve noise cancellation up to specific orders in time-dependent perturbation theory. Albeit rather unrealistic, this operating scenario has served as the foundation for the development of numerous DD protocols~\cite{viola1998dynamicalsuppression,zanardiSymmetrizingEvolutions1999,viola1999dynamicaldecoupling,khodjasteh2005faulttolerantquantum,uhrig2007keepingquantum}, many of which have performed well on modern quantum devices~\cite{pokharel2018demonstrationfidelity, ezzell2023dynamicaldecoupling}. Moreover, they have motivated sequence constructions that combat experimentally relevant non-idealities in the control, e.g., errors in the pulse rotation angle~\cite{wang2012dd, genov2017arbitrarilyaccurate, souza2012robust}. When accounting for the finite-amplitude nature of control pulses, referred to as the bounded-control setting, errors induced by the finite pulses have also been a topic of interest for many years~\cite{viola2003robustdynamical,pryadko2008shaping, khodjasteh2009dcgs,gregoryquiroz2013optimizeddynamical}.

Numerous foundational DD studies center on single-qubit error suppression, however, the technique is by no means limited to this domain. Multi-qubit protocols based on Hadamard matrices have been shown to offer suppression of internal coupling in quantum registers in the ideal-pulse limit~\cite{jones1999efficient,stollsteimer2001dd,leung2002simulation}. Extensions leveraging orthogonal arrays have also been identified~\cite{rotteler2006equivalence}, with subsequent works drawing on balanced-cycle and Eulerian orthogonal arrays in the bounded-control setting~\cite{wocjan2006efficientdecoupling, bookatz2016improvedboundedstrength}.

In recent years, there has been a resurgence in the investigation of multi-qubit DD for parasitic internal coupling. This work has been driven in part by the presence of spatially correlated $ZZ$ errors observed in fixed-coupler superconducting qubit architectures~\cite{krantz2019quantumengineers,fors2024comprehensiveexplanation}. By exploiting the topology of the qubit array, modern DD protocols have sought to address $ZZ$ errors while improving on pulse overheads. For example, in the ideal-pulse limit, efforts have centered on reducing the total number of pulses~\cite{brown2025efficientchromatic}.

Pulse staggering has emerged as a technique for suppressing internal coupling in the bounded-control regime. Here, pulse are applied asynchronously between neighboring qubits situated on a 2-colorable topology. The staggering results in a cancellation of all $ZZ$ errors, including those induced by the finite-amplitude nature of the control. Importantly, while the total protocol execution time increases by a factor of 2, the number of pulses remains unaltered. Such techniques have been referred to as crosstalk-robust DD~\cite{zhou2023crctrl} as well as syncopated DD~\cite{evert2024syncopateddynamical}. Analytical understanding of pulse staggering in the bounded-control setting has been extremely limited~\cite{zhou2023crctrl}, and it is unclear whether this approach extends to broader classes of DD protocols.

In this work, we show that pulse staggering indeed extends to a wide variety of DD sequences and sequence compositions. In particular, we find this approach to be broadly applicable to sequences of equidistant $\pi$-pulses. The protocol does not rely on using the same sequence on each qubit within the qubit array, enabling a versatile approach for composing DD sequences while maintaining $ZZ$ robustness. We motivate our findings via symmetry and orthogonality arguments based on the crosstalk suppression condition identified in Ref.~\cite{zhou2023crctrl}. 

Numerical validation is accompanied by a series of demonstrations performed on the IBM Quantum Platform (IBMQP). Using fixed-coupler architectures, we apply our approach to numerous well-known DD sequences, and study the impact of pulse staggering on collections of up to 20 qubits. We observe approximately $3\times$ to $10\times$ improvement in the characteristic time of fidelity decay through staggering. Moreover, it is shown that the protocol is amenable to inter-pulse padding of idle periods. These findings speak to the utility of pulse staggering in DD-protected algorithm implementations and integrated error management strategies, where DD is used to protect idling qubits. 

Lastly, we use the protocol to assess alterations in hardware design. We examine the performance advantages of $ZZ$-robust DD on IBMQP tunable-coupler devices designed to mitigate $ZZ$ errors at the hardware layer~\cite{stehlik2021tunable}. Our findings confirm that such devices attain substantial suppression of $ZZ$ errors as desired. Despite this architectural advantage and faster gate operations, our demonstrations indicate that fixed-coupler architectures equipped with crosstalk-robust DD (CR-DD) can outperform DD-protected evolution on tunable-coupler systems. The results of our study broaden the scope of high-performing DD and its utility on modern quantum computing systems.

The structure of this paper is as follows. In Sec.~\ref{sec:dd-bg}, we briefly summarize background information related to the suppression conditions for DD. In Sec.~\ref{sec:dd-sup-cond}, we lay out the specifics of the error model that our protocol is founded on and define a first-order suppression condition for 1-local and 2-local noise. In Sec.~\ref{sec:dd-detail} we detail DD sequences from prior work that we use in our numerical and experimental studies. In Sec.~\ref{sec:eng-suppression}, we discuss protocols for $ZZ$ noise suppression in both the ideal-pulse and bounded-control settings. Via our numerical study, discussed in Sec.~\ref{sec:numerical-study}, we provide evidence for the claims and intuition behind pulse staggering. Finally, we introduce our experimental methods and comparisons in Sec.~\ref{sec:experiments} and make closing remarks in Sec.~\ref{sec:conclusion}.

\section{\label{sec:dd-bg}Dynamical Decoupling Background}
\subsection{Effective Error Dynamics}
A common setting for DD is that of an open quantum system described by the Hamiltonian
\begin{equation}\label{eq:Htot}
H(t) = H_C(t) + H_{\text{err}}.
\end{equation}
$H_C(t)$ is the control Hamiltonian that acts solely on the system Hilbert space $\mathcal{H}_S$ and defines the specifications of the DD sequence over a cycle duration $\tau_c$. In contrast, $H_{\text{err}}$ mediates errors generated by unwanted pure system evolution and interactions within the joint system and environment Hilbert space $\mathcal{H}_S \otimes \mathcal{H}_E$. 

Formally, the dynamical evolution of Eq.~\eqref{eq:Htot} can be described by $U(t)=\mathcal{T}_+ \exp\left(-i \int^t_0 H(s) ds\right)$, where $\mathcal{T}_+$ denotes the time-ordering operator. However, the dynamics can be examined via time-dependent perturbation theory under the assumption of strong control, or equivalently, weak noise~\cite{khodjasteh2007performancedeterministic}. Mathematically, this assumption is satisfied if $\|H_C(t)\|\gg \|H_{\text{err}}\|$, where $\|A\|$ denotes the sup-operator norm or largest singular value of an operator $A$.

The perturbative analysis begins by moving into the rotating frame with respect to the control and defining the Hamiltonian $\tilde{H}_{\text{err}}(t)=U^{\dagger}_C(t) H_{\text{err}}U_C(t)$. As a result, the dynamics can be partitioned as $U(t)=U_C(t) \tilde{U}_{\text{err}}(t)$, where $U_C(t)=\mathcal{T}_+ \exp\left(-i \int^t_0 H_C(s) ds\right)$ dictates the pure control evolution and the error dynamics are described by $\tilde{U}_{\text{err}}(t)=\mathcal{T}_+ \exp\left(-i \int^t_0
\tilde{H}_{\text{err}}(s) ds\right)$. Utilizing the Magnus expansion~\cite{moan1999existence}, the error dynamics can be expressed as 
\begin{equation}
    \tilde{U}_{\text{err}}(t) = \exp\left(\sum^{\infty}_{n=1} (-i)^n\Omega^{(n)}(t)\right)
\end{equation}
with the anti-Hermitian operators $\Omega^{(n)}(t)$ representing the $n$th Magnus term. The first two orders of the expansion are given by
\begin{eqnarray}
    \Omega^{(1)}(t) &=& \int^{t}_0 \tilde{H}_{\text{err}}(s)\, ds\nonumber\\
    \Omega^{(2)}(t) &=& \frac{1}{2}\int^{t}_0 ds_1 \int^{s_1}_0 ds_2\, [\tilde{H}_{\text{err}}(s_1), \tilde{H}_{\text{err}}(s_2)]\nonumber,
\end{eqnarray}
while higher orders can be constructed recursively as a sum of $(n-1)$-fold commutators.

Within this framework, the effectiveness of DD can be quantified by its ability to suppress successive orders of Magnus terms. In particular, a DD sequence is said to achieve $k$th order decoupling if $\Omega^{(n\leq k)}(\tau_c)=0$. Noting that DD is ideally identity evolution, $U_C(\tau_c)=I$, the resulting evolution of Eq.~\eqref{eq:Htot} can be expressed as
\begin{eqnarray}
    U(\tau_c) &=& \exp\left({\sum^{\infty}_{n=k+1}(-i)^n\Omega^{(n)}(\tau_c)}\right) \nonumber\\
    &\eqsim& I + \mathcal{O}(\|H_{\text{err}}\|^{k+1}\tau^{k+1}_c).
\end{eqnarray}

\subsection{Control Matrix Representation}
\label{subsec:ctrl-matr}
An alternative perspective on DD is afforded by the so-called control matrix. Commonly associated with the filter function formalism~\cite{kofman2001universal, cywinski2008ff, green2012highorder, green2013arbitrary, pazsilva2019qns}, the control matrix representation relates the control dynamics to the error Hamiltonian through time-dependent modulation functions. An approach to deriving the control matrix representation in general settings is to define $H_{\text{err}}={\sum_\alpha S_\alpha \otimes B_\alpha}$. The $S_\alpha$ form a basis of traceless system operators with their orthogonality defined via $\text{Tr}[S_\alpha S_\beta]/N=\delta_{\alpha,\beta}$. The dimension of the Hilbert space is $N=\dim(\mathcal{H}_S)$ and $\delta_{\alpha,\beta}$ is the Kronecker delta function. An example of such a basis is the $n$-qubit Pauli basis, where $N=2^n$. Environment operators $B_\alpha$ are bounded and defined on the environment Hilbert space. When $B_\alpha\propto I$, the errors are due to pure system imperfections and designate system-environment interactions otherwise. 

The basis expansion of the error Hamiltonian enables the Magnus terms to be written in an alternative form. In particular, the first-order Magnus term can be expressed as
\begin{equation}
    \Omega^{(1)}(\tau_c) = \sum_{\alpha,\beta} \chi^{\alpha\beta}(\tau_c) S_\beta \otimes B_\alpha,
\end{equation}
where the error matrix
\begin{equation}\label{eq:gen-err-matr}
    \chi^{\alpha\beta}(\tau_c) = \int^{\tau_c}_0 R^{\alpha\beta}(t) dt.
\end{equation}
The control matrix
\begin{equation}\label{eq:ctrl-matr}
    R^{\alpha\beta}(t) = \frac{1}{N}\text{Tr}[U^\dagger_C(t) S_\alpha U_C(t) S_\beta]
\end{equation}
defines the time-dependent projections of the rotated error Hamiltonian operators on the system basis. Equivalent expressions can be derived for higher order Magnus terms as well.

The desired effect of DD in the control matrix representation is to provide suppression of the error matrix for each Magnus term. Thus, first-order suppression is achieved when~\cite{zhou2023crctrl}
\begin{equation}\label{eq:gen-supp-cond}
    \chi^{\alpha\beta}(\tau_c)=0 \quad \forall \alpha,\beta.
\end{equation}
This is analogous to imposing conditions on $R^{\alpha\beta}(t)$ such that cancellation occurs over the time interval of the DD cycle. Below, we will utilize this condition to define conditions under which DD sequences can be derived. Furthermore, it will be employed to provide intuition for our proposed approach to DD design.

\section{\label{sec:dd-sup-cond}DD Suppression Condition for 1-Local Errors and \texorpdfstring{$ZZ$}{ZZ} Crosstalk}
\subsection{System Model}
\label{subsec:sys-model}
Here, we will focus on a system of $n$ qubits whose topology is defined by a graph $G=(V,E)$. The qubits reside on the vertices defined by the set $V$ with $|V|=n$, where interaction between qubits $v$ and $w$ in $V$ is represented by an edge $e=(v, w)$ in $E$. This graph of qubit interactions is not required to be isometric to the hardware topology. We assume the DD sequences are composed of single-qubit gates. The gates are generated by the control Hamiltonian
\begin{equation}
    H_C(t)=\sum_{v\in V} \frac{\omega_v(t)}{2}\left[\sigma^X_v \cos\phi_v(t) + \sigma^Y_v\sin\phi_v(t)\right]
    \label{eq:hc-sq}
\end{equation}
where the amplitude and phase, $\omega_v(t)$ and $\phi_vt)$, respectively, constitute the control degrees of freedom for the $i$th qubit. As such, the DD sequences are not limited to single-axis rotations. Note that $\{\sigma^\mu_v\}$, with $\mu=X,Y,Z$ denotes the Pauli operators on qubit $v$.

We consider an error Hamiltonian of the form 
\begin{equation}
    H_{\text{err}} = \sum_{v\in V} \sum_{\mu=X,Y,Z} \sigma^{\mu}_v\otimes B^\mu_v + \sum_{e \in E} \sigma^Z_v\sigma^Z_w\otimes B^{ZZ}_{e}
    \label{eq:H-err}
    \end{equation}
The first term defines single-qubit (1-local) errors, while the second term includes two-qubit (2-local) $ZZ$ errors. Relative to the topology of the qubits, note that the 1-local terms are determined by the number of vertices, while the 2-local errors are determined by the edges. This Hamiltonian generically captures all 1-local and $ZZ$ interactions between qubits, whether they be purely due to unwanted internal system dynamics or interactions with the environment. For example, by setting $B^{ZZ}_e=J^{ZZ}_e I$, we obtain parasitic spatially correlated, quantum crosstalk interactions determined by the coupling strength $J^{ZZ}_e$~\cite{zhou2023crctrl}.

\subsection{First-Order Suppression Condition}
\label{subsec:first-order-cond}
We derive a first-order suppression condition using the control matrix representation. We follow the approach in Sec.~\ref{subsec:ctrl-matr} by moving into the rotating frame with respect to the control evolution operator. While the 1-local error Hamiltonian follows directly from Eq.~\eqref{eq:gen-err-matr}, the 2-local terms include control matrices that involve interactions between qubits. However, since the control is local, the control unitaries factorize as $U_C(t)=\bigotimes_{v\in V}U_{C,v}(t)$. The resulting error matrix for the 2-local term is therefore composed of a product of single-qubit control matrices. 

Specifically, the first-order Magnus term can be written as 
\begin{equation}
    \Omega^{(1)}(\tau_c) =\sum_{v\in V}\Omega^{(1)}_{1,v}(\tau_c) + \sum_{e\in E}\Omega^{(1)}_{2,e}(\tau_c),
    \label{eq:first-order-magnus}
\end{equation}
where
\begin{eqnarray}
    \Omega^{(1)}_{1,v}(\tau_c) &=& \sum_{\mu,\alpha=X,Y,Z} \chi^{\mu\alpha}_{1,v}(\tau_c)\, \sigma^{\alpha}_v \otimes B^\mu_v \label{eq:omega1a}\\
    \Omega^{(1)}_{2,e}(\tau_c) &=& \sum_{\alpha,\beta=X,Y,Z} \chi^{ZZ\alpha\beta}_{2,e}(\tau_c)\, \sigma^{\alpha}_v\sigma^{\beta}_w \otimes B^{ZZ}_{e}.\quad\quad 
\end{eqnarray}
The 1-local error matrix is given by
\begin{equation}
    \chi^{\mu\alpha}_{1,v}(\tau_c) = \int^{\tau_c}_0 R^{\mu\alpha}_v(t) dt,
    \label{eq:1-local-err-matr}
\end{equation}
where $R^{\mu\alpha}_v(t)=\text{Tr}[U^{\dagger}_{C,v}(t)\sigma^\mu_v U_{C,v}(t) \sigma^{\alpha}_v]/2$, and thus determined solely by the individual control sequences. In contrast, the 2-local error matrix
\begin{equation}\label{eq:2-local-cond}
    \chi^{ZZ\alpha\beta}_{2,e}(\tau_c) = \int^{\tau_c}_0 R^{Z\alpha}_v(t)R^{Z\beta}_w(t) dt
\end{equation}
is dictated by the control applied to qubits $v$ and $w$. This dependence poses challenges as, from a DD perspective, implies that simultaneous sequence design is necessary in order to address 2-local errors.

Based on the 1- and 2-local error matrices, we define the first-order suppression conditions as
\begin{eqnarray}
    \chi^{\mu\alpha}_{1,v}(\tau_c) &=& 0, \forall v,\mu,\alpha \label{eq:dd-supp-cond1}\\
    \chi^{ZZ\alpha\beta}_{2,e}(\tau_c) &=& 0, \forall e,\alpha,\beta. \label{eq:dd-supp-cond2}
\end{eqnarray}
Similar to Eq.~\eqref{eq:gen-supp-cond}, the 1-local suppression condition can be achieved by engineering the control matrix components for individual qubits. In the case of the 2-local condition, the product of control matrices must be properly engineered. While seemingly non-trivial to satisfy, we show that both can be achieved in the context of DD in a straightforward manner—even in the case of bounded controls.

\section{\label{sec:dd-detail} DD Protocols for 1-Local Errors}
In this section, we review DD protocols pertinent to this study. We select some of the best-performing sequences tested on several quantum devices offered by the IBMQP~\cite{ezzell2023dynamicaldecoupling}. These sequences are designed to address 1-local errors. However, in Sec.~\ref{sec:eng-suppression}, we will promote them to $ZZ$ suppressing protocols using a pulse staggering approach. We will then explore their efficacy numerically and demonstrate them on quantum hardware.

\subsection{General DD Sequence Construction}
DD evolution typically takes the form
\begin{equation}\label{eq:general-dd}
    U_{\rm DD}(\tau_c) = P_K f_{K} P_{K-1} f_{K-1} \cdots P_{1} f_{1},
\end{equation}
such that $K$ pulses are applied successively with periods of free evolution interleaved between them. The pulses $P_j$ are determined by the control Hamiltonian $H_C(t)$ and ideally are both error-free and applied instantaneously; this is the the so-called bang-bang limit~\cite{viola1998dynamicalsuppression}. However, in practice, the pulses are prone to errors and bounded in amplitude. As a result, during the pulse duration $\tau_p$, the error Hamiltonian can contribute to the overall pulse dynamics~\cite{lidar2014review}. Free evolution periods defined by $f_{j}=e^{-i \tau_j H_{\text{err}}}$ occur over a specified time duration $\tau_j$. This interpulse time denotes periods during which the system evolves according to its internal dynamics, which are assumed to be solely generated by the error Hamiltonian. The total DD evolution time is given by $\tau_c=\sum_j \tau_j + K\tau_p$. 

Below, we discuss sequences that satisfy Eq.~\eqref{eq:dd-supp-cond1} in the ideal- and imperfect-pulse regimes. Sequences of $\pi$-pulses, i.e., $\int^{\tau_c}_0 \omega(t)\, dt=\pi$ will be the focus, and will be denoted by $(\pi)_\phi$, where $\phi$ denotes the phase during the pulse [see Eq.~\eqref{eq:hc-sq}]. Furthermore, we focus on sequences with fixed inter-pulse delays $\tau_d$, i.e., $\tau_j=\tau_d$  $\forall j$. Thus, our discussion will not include sequences, such as the Uhrig DD (UDD) sequences, known to achieve efficient decoupling in the ideal, instantaneous pulse regime by optimizing the inter-pulse spacing~\cite{uhrig2007keepingquantum}.

\subsection{XY4}
The XY4 universal decoupling sequence~\cite{maudsley1986modifiedcarrpurcellmeiboomgill} is a well-known DD protocol that affords first-order suppression in the ideal-pulse limit. Specifically, it enables suppression of 1-local errors for generic linear system-bath coupling between a single qubit and quantum environment. Furthermore, XY4 serves as the foundation for higher-order protection DD schemes, such as concatenated DD (CDD)~\cite{khodjasteh2005faulttolerantquantum,khodjasteh2007performancedeterministic}. 

The XY4 sequence is typically defined as
\begin{eqnarray}
\text{XY4} &\equiv &Y-f_d-X-f_d-Y-f_d-X-f_d\\
&= &\prod_{\phi \in (0, \frac{\pi}{2}, 0, \frac{\pi}{2})} (\pi)_\phi f_d,
\end{eqnarray}
where alternating $\pi$-pulse are applied along the $X$ and $Y$ axes of the single-qubit Bloch sphere and $\Pi$ indicates left-multiplication of operators as in Eq.~\eqref{eq:general-dd}. The sequence is constructed by symmetrizing the error Hamiltonian under the action of a decoupling group~\cite{zanardiSymmetrizingEvolutions1999,viola1999dynamicaldecoupling}. This decoupling group is generated by the single-qubit Pauli operators. Symmetrization guarantees first-order suppression in the limit of ideal, instantaneous pulses.

\subsection{Knill Dynamical Decoupling}
The Knill DD protocol (KDD)~\cite{souza2011robustdynamical} is a twenty-pulse sequence that builds on XY4. It is obtained by substituting 5-pulse composite pulse sequences that apply an effective X or Y rotation that is robust to flip-angle errors (i.e., over/under-rotation errors) and off-resonance errors. KDD can be expressed as
\begin{eqnarray}
\text{KDD} &\equiv & K_Y - K_X - K_Y - K_X \\
K_X &\equiv &\prod_{\phi \in (\frac{\pi}{6}, 0, \frac{\pi}{2}, 0, \frac{\pi}{6})} (\pi)_\phi f_d\\
K_Y &\equiv &\prod_{\phi \in (\frac{2\pi}{3}, \frac{\pi}{2}, \pi, \frac{\pi}{2}, \frac{2\pi}{3})} (\pi)_\phi f_d,
\end{eqnarray}
where component $\pi$-pulses of the composite pulses are interleaved with delays inside the product expressions.

\subsection{Universally Robust Dynamical Decoupling}
Universally robust DD (URDD)~\cite{genov2017arbitrarilyaccurate} refers to a parametrized family of DD sequences. Each sequence is defined by $\pi$-pulses whose phases are carefully chosen to provide robustness against flip-angle errors and static detuning errors. The pulse phases are constructed recursively based on suppression conditions identified from a perturbative analysis of the transition probability.

Experimental investigations of the UR sequences have shown promising results. In Ref.~\cite{ezzell2023dynamicaldecoupling}, it was found that the performance of the UR family on two earlier IBM superconducting qubit devices, \bogota{} and \jakarta{}, was saturated at a UR sequence length of ten to twenty pulses depending on the connectivity graph. For the symmetrized UR10 sequence, we have
\begin{eqnarray}
\Phi &=& \frac{\pi}{5}(0, 4, 2, 4, 0, 0, 4, 2, 4, 0)\\
\text{UR10} &\equiv& \prod_{\phi \in \Phi} (\pi)_{\phi} f_d.
\end{eqnarray}
This sequence will serve as a testing ground for our $ZZ$ suppression protocol applied to the UR family. However, we note that any order UR can in principle be utilized.

\subsection{Robust Genetic Algorithm Dynamical Decoupling}
The robust genetic algorithm (RGA) sequences are a family of DD sequences that were discovered through numerical search via genetic algorithms~\cite{gregoryquiroz2013optimizeddynamical}. In this work, we specifically focus on RGA$_{64c}$ sequence, which we form by a virtual concatenation~\cite{alvarez2012iterativerotation} of an RGA$_{8c}$ sequence, or equivalently, an Eulerian DD (EDD) sequence, with another RGA$_{8c}$ sequence. 

The EDD sequence extends the group averaging approach to the bounded-control setting by applying the pulses according to an Eulerian path through the Cayley graph specified by a decoupling group. EDD achieves first-order decoupling using an eight-pulse palindromic sequence that builds on XY4~\cite{viola2003robustdynamical}:
\begin{eqnarray}
\text{EDD} &\equiv &\prod_{\phi \in (0, \frac{\pi}{2}, 0, \frac{\pi}{2}, \frac{\pi}{2}, 0, \frac{\pi}{2}, 0)} (\pi)_\phi f_d.
\end{eqnarray}
Key to its design, EDD is capable of outperforming XY4 if the pulse duration is sufficiently larger than the inter-pulse delay. Moreover, the EDD concept can be extended to other sequences as well~\cite{wocjan2006efficientdecoupling}, and it serves as the basis for dynamical corrected gates (DCGs)~\cite{khodjasteh2009dynamicallyerrorcorrected}. 

Utilizing EDD in the RGA construction yields a 64-pulse DD cycle robust to pulse-width errors. More concretely, the concatenation is described as
\begin{eqnarray}
    \text{RGA}_{64c} &=& \text{RGA}_{8c}[\text{RGA}_{8c}]\nonumber\\
    &=& \prod_{\phi \in (0, \frac{\pi}{2}, 0, \frac{\pi}{2}, \frac{\pi}{2}, 0, \frac{\pi}{2}, 0)} (\pi)_\phi \left(\text{EDD}\right),
\end{eqnarray}
where the free evolution periods within EDD are replaced with an additional layer of EDD.

\section{\label{sec:eng-suppression}Engineering \texorpdfstring{$ZZ$}{ZZ} Suppression by Selective Pulse Timing}
Interplay between the sequences poses a challenge for satisfying the 2-local suppression condition. Sequences applied to qubits with a common edge must be carefully chosen to achieve the desired suppression. Here, we show how to suppress $ZZ$ interactions by shifting the location of the pulses in a DD sequence. The approach is useful for 2-colorable qubit topologies. While we briefly discuss approaches for achieving suppression in the ideal-pulse limit, we focus on practical implementation in which the pulses are bounded in amplitude and possess finite duration. Importantly, our protocol is not limited to a specific DD sequence, but rather extends to a family of sequences commonly used on modern quantum hardware. The sequences presented in Sec.~\ref{sec:dd-detail} are examples that fall within this family.

First, we note the definition of graph coloring since it will be an important part of our sequence design. For a graph $G$, a coloring is an assignment of natural numbers (colors) to $V$ such that no two vertices that are connected by an edge are assigned the same color. A $k$-coloring of $G$ is a valid coloring of $G$ with at most $k$ colors. Vertices assigned to the same color are said to be in the same color class. A common convention for 2-colorable graphs is to label the color classes of the vertices red ($R$) and blue ($B$). We will focus on 2-colorable graphs given their relevance to existing quantum hardware topologies~\cite{murali2019full, chamberland2020}. 

Obtaining error suppression for a 2-colorable graph can be reduced to engineering sequences that satisfy Eq.~\eqref{eq:dd-supp-cond2}. An approach is to start with the control matrices and demand certain symmetries be enforced on the individual components. Sequences are then determined based on the selected control matrices. In Ref.~\cite{paz2016dynamical}, the symmetry argument proved useful in identifying DD sequences for an $n$-qubit local dephasing model. Two types of (anti)symmetry were identified.

\emph{Displacement (anti)symmetry} refers to the repetition of a function in time,
\begin{equation}\label{eq:disp-symm}
    R^{\mu\alpha}_v(t+\tau_c/2)=\pm R^{\mu\alpha}_v(t),\ t\in [0,\tau_c/2].
\end{equation}
\emph{Mirror (anti)symmetry} refers to the reversal of a function in time,
\begin{equation}\label{eq:mirror-symm}
    R^{\mu\alpha}_v(\tau_c-t) = \pm R^{\mu\alpha}_v(t),\ t\in [0,\tau_c/2].
\end{equation}
Displacement- and mirror anti-symmetry are both obtained by introducing the sign flip in half of the domain. We show how this language can be extended to aid the design of sequences robust to 2-local $ZZ$ errors as well.

\subsection{Ideal Pulses}
We illustrate the utility of symmetrization by first considering the bang-bang setting. Here, a strategy is to choose sequences that give an antisymmetric product of control matrix components. This can be achieved by selecting the pulse timing such that one color's sequence contains time-asymmetric building blocks (each pulse is preceded or followed by a pulse-length delay) and the other contains time-symmetric building blocks (each pulse is surrounded by half-pulse-length delays). The DD-to-color assignment and the correspondence of color to the two sequences are arbitrary.

Here, we choose $R$ to denote the sequence with asymmetric delay of $\tau_d$ and $B$ the sequence with symmetric delay of $\tau_d/2$: 
\begin{eqnarray}
    \text{DD}_R &=& P_K f_d P_{K-1} f_d \cdots P_2 f_d P_1 f_d\\
    \text{DD}_B &=& \sqrt{f_d}P_K f_d P_{K-1} \cdots f_d P_2 f_d P_1 \sqrt{f_d}.
\end{eqnarray}
An example is that of $K=4$ with ${P_1=P_3=X}$ and ${P_2=P_4=Y}$. This leads to the XY4 universal decoupling sequence with asymmetric and symmetric delay timing, respectively.

\subsection{Bounded Control}
In the case of bounded control, finite pulse duration spoils the suppression conditions provided by the symmetry-asymmetry sequence pair. It is possible to regain the suppression, however, via an alternative approach. We present the method first by example. Consider an XY4 protocol defined by the sequences
\begin{equation}\label{eq:cr-xy4}
\begin{aligned}
    \text{DD}_R &= Y f_p X f_p Y f_p X f_p  \\
    \text{DD}_B &= f_p Y f_p X f_p Y f_p X,
\end{aligned}
\end{equation}
where the delays of pulse duration $\tau_p$ offset the pulse execution. Identified in Ref.~\cite{zhou2023crctrl}, this sequence pair was shown to satisfy Eq.~\eqref{eq:dd-supp-cond2} for bounded control and its efficacy was explored through demonstrations therein.

The main result of this work is that one can use this timing strategy with all single-qubit DD sequences composed of $\pi$-pulses separated by equal delays. Specifically, consider a single-qubit DD sequence with $\pi$-pulses $(\pi)_{\phi_j}$ of duration $\tau_p$, where $\phi_j$ are chosen from the ordered set $\Phi$.  We find that the family of 2-color sequences given by
\begin{equation}\label{eq:cr-dd}
\begin{aligned}
    \text{DD}_R &= \prod_{\phi_j \in \Phi_R} (\pi)_{\phi_j} f_p \\
    \text{DD}_B &= \prod_{\phi_j \in \Phi_B} f_p (\pi)_{\phi_j}.
\end{aligned}
\end{equation}
are sufficient for suppressing all effective noise components arising from $ZZ$ errors while retaining the original sequences' robustness to single-qubit errors. We will refer to the protocol in Eq.~\eqref{eq:cr-dd} as $\text{CR-}(\text{DD}_R, \text{DD}_B)$ below when $\Phi_R$ and $\Phi_B$ are unique and as $\text{CR-DD}$ when they are the same, where the name of the respective DD sequence will be substituted for ``DD." Note that $\Phi_R$ and $\Phi_B$ do not need to generate the same sequence of pulses, however, the sizes of the lists should be equal, i.e., $L=|\Phi_R|=|\Phi_B|$. As such, the total sequence duration is given by $\tau_c=2L\tau_p$. Below, we show that sequences of differing lengths can be combined through sequence repetition.

Moreover, we find that the robustness is preserved when equally padding any amount of additional delay $\tau_d$  around the pulses either symmetrically,
\begin{equation}\label{eq:sym-pad-cr-dd}
\begin{matrix}
    \text{DD}_R = \prod_{\phi_j \in \Phi_R} &\left[ \sqrt{f_d} \right.  &(\pi)_{\phi_j} &f_d  &f_p  &\left. \sqrt{f_d} \right] \\
    \text{DD}_B = \prod_{\phi_j \in \Phi_B} &\left[ \sqrt{f_d} \right. &f_p  &f_d  &(\pi)_{\phi_j} &\left. \sqrt{f_d} \right],\\
\end{matrix}
\end{equation}
or asymmetrically,
\begin{equation}\label{eq:asym-pad-cr-dd}
    \begin{matrix}
        \text{DD}_R = \prod_{\phi_j \in \Phi_R} &\left[(\pi)_{\phi_j} \right. & f_d & f_p & \left. f_d \right] \\
        \text{DD}_B = \prod_{\phi_j \in \Phi_B} &\left[ f_p\right. & f_d & (\pi)_{\phi_j} & \left. f_d \right],\\
    \end{matrix}
\end{equation}
where the choice of leading or trailing with the additional delay in the asymmetric case is arbitrary. With the robustness intact, we retain analytic elimination of spatially-correlated $ZZ$ noise from the suppression condition. Furthermore, we gain the freedom to explore the tradespace between pulse number and sequence duration to improve sequence performance. Both padded sequences have a duration $\tau_c=2L(\tau_p + \tau_d)$. When $\Phi_A=\Phi_B$ and $\tau_d = (k-1) \tau_p$, for $k\ge1, k\in\mathbb{N}$, we will refer to Eqs.~\eqref{eq:sym-pad-cr-dd}~and~\eqref{eq:asym-pad-cr-dd} as $\text{CR-DD-}k_S$ and $\text{CR-DD-}k_A$, respectively.  In this case, we have $\tau_c = 2kL\tau_p$ for both padded CR sequences.

For convenience, we can also express (non-crosstalk-robust) simultaneous DD with bounded controls in a similar fashion. We define simultaneous DD in terms of a single DD sequence given by $\Phi$ with $|\Phi|=L$ and a {delay $\tau_d$,}
\begin{equation}\label{eq:sim-dd}
\text{SIM-DD} = \prod_{\phi_j \in \Phi} [f_d\ (\pi)_{\phi_j}].
\end{equation}
This cycle will have duration $\tau_c = L(\tau_p + \tau_d)$. When $\tau_d=(k-1)\tau_p$ we will refer to Eq.~\eqref{eq:sim-dd} as $\text{SIM-DD-}k$ which has duration $\tau_c=kL\tau_p$. Note that we omit the suffix when $k=1$ and there is no additional padding in a sequence.

\section{\label{sec:numerical-study}Numerical study of crosstalk robustness}
Here, we provide intuition for pulse staggering error suppression through a numerical investigation. We study the control matrices and suppression condition [Eq.~\eqref{eq:2-local-cond}] dynamics as a function of time for a variety of DD protocols. Specific control matrix components are used to illustrate the effect of pulse staggering and characterize its impact on $ZZ$ crosstalk suppression. We identify key features that enable the technique to extend to a wide variety of sequence compositions. 

Below, two types of sequence compositions are discussed. The first we denote as homogeneous sequence composition. We use this label to refer to the case where the sequences being composed are of the same type. For example, the case when $\Phi_R=\Phi_B$ in Eq.~\eqref{eq:cr-dd}. In addition, we consider heterogeneous sequence composition in which the sequences are distinct (i.e., $\Phi_R\neq\Phi_B$). We find pulse staggering offers suppression of $ZZ$ errors in both cases, in part due to the same characteristic suppression properties. In each example below, we consider bounded control described by Derivative Removal by Adiabatic Gate (DRAG) pulses~\cite{motzoi2009drag}, however, we note that the results can be extended to any finite-width pulse profile.

\subsection{Homogeneous Sequence Composition}
\begin{figure*}
    \centering
    \includegraphics[width=\textwidth]{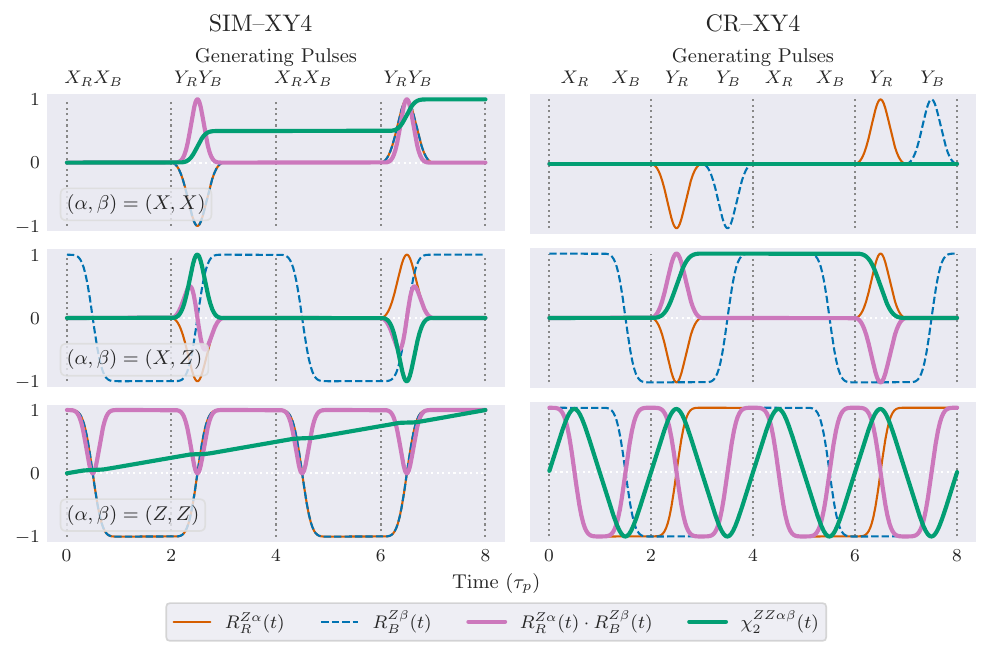}
    \caption{A side-by-side comparison of selected pairs of elements from the control matrices $R^{Z\alpha}_R(t)$ and $R^{Z\beta}_B(t)$ of SIM-XY4 (left) and CR-XY4 (right) with bounded control (DRAG pulses) through one cycle. The corresponding product of the control matrix elements and the elements of $\chi_2^{ZZ\alpha\beta}(t)$ that the pairs generate are also shown. SIM-XY4 exhibits error accumulation in certain cases, e.g., $(\alpha,\beta)=(X,X), (Z,Z)$, whereas CR-XY4 provides complete cancellation.}
    \label{fig:integrator-xy4}
\end{figure*}

\begin{figure*}
    \centering
    \includegraphics[width=\textwidth]{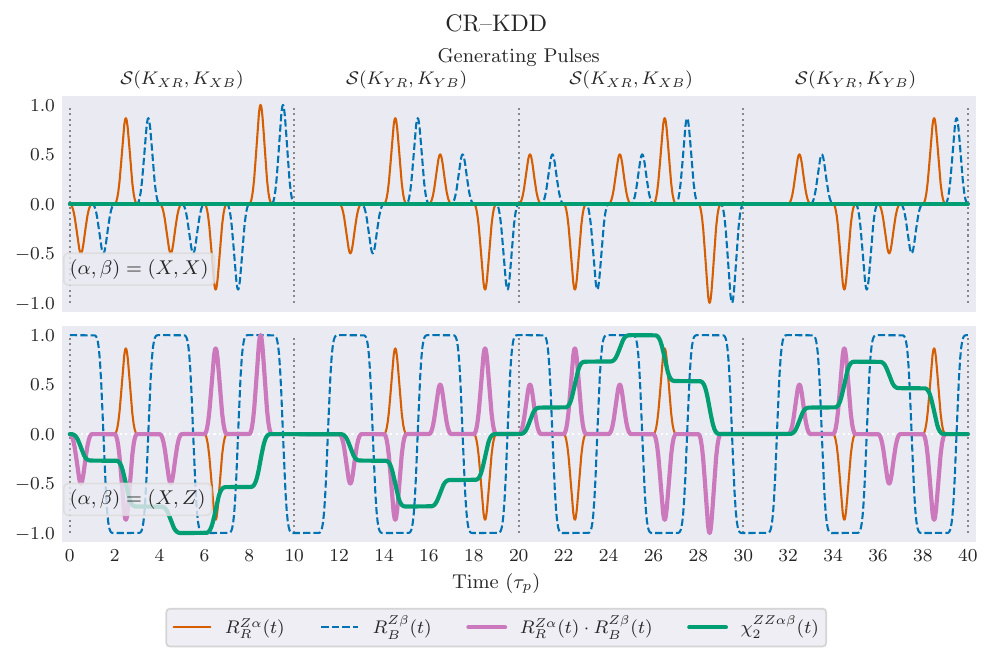}
    \caption{Selected pairs of elements from the control matrices of CR-KDD with bounded control (DRAG pulses) through one cycle, their product, and the 2-local error matrix, $\chi_2^{ZZ\alpha\beta}(t)$. Similar to CR-XY4, pulse staggering affords the necessary symmetry and orthogonality to suppress $ZZ$ errors at the end of the sequence.}
    \label{fig:integrator-kdd}
\end{figure*}

The discussion on homogeneous sequence composition will focus on XY4 and KDD. The former provides a simple example to build intuition, while the latter exhibits a more complex illustration of pulse staggering noise suppression. Numerical results for both sequences are summarized in Figs.~\ref{fig:integrator-xy4} and \ref{fig:integrator-kdd}.

We utilize XY4 to convey the distinction between simultaneous and staggered pulse placement. In Fig.~\ref{fig:integrator-xy4}, we show the control matrices $R^{Z\alpha}_R(t)$ and $R^{Z\beta}_B(t)$ as a function of time for $(\alpha,\beta)=(X,X), (X,Z), (Z,Z)$. Simultaneous XY4 (SIM-XY4) and staggered, crosstalk-robust XY4 (CR-XY4) are shown in the left and right columns, respectively. The control matrix components are accompanied by their product and the error matrix components $\chi^{ZZ\alpha\beta}_2(t)$. (We omit specifying the edge subscript here because every edge in a bipartite graph connects a member of $R$ to a member of $B$.) The top of each column includes the pulse locations for further clarity.

We focus on these three control matrix components as they capture three distinct types of dynamics. The first, in which $(\alpha,\beta)=(X,X)$, stems from dynamics generated by finite-width pulses. (In the bang-bang limit, the control matrix is given solely by $R^{ZZ}$.) The exact dynamics of the $(X,X)$ control matrix component are dictated by the pulse profile. In Fig.~\ref{fig:integrator-xy4}, the distinction between simultaneous and staggered pulses with respect to $(\alpha,\beta)=(X,X)$ is shown in the first row. The control matrices possess overlapping contributions in the simultaneous case, while remaining orthogonal in the crosstalk-robust implementation. As a result, the subsequent error matrix component for SIM-XY4 and CR-XY4 are non-zero and zero, respectively. Similar characteristics are observed for $(\alpha,\beta)=(Y,Y)$. In contrast, cross-terms purely generated by pulse errors, e.g., $(\alpha,\beta)=(X,Y)$ cancel for both sequence types. This follows directly from the fact that $X$ and $Y$ pulses are never applied simultaneously in either sequence. At most, simultaneous operation is limited to solely $X$ or $Y$ pulses, as in the case of SIM-XY4.

Next, consider the case of $(\alpha,\beta)=(Z,Z)$. These components correspond to $ZZ$ errors that are preserved at time $t$. As shown in the third row of Fig.~\ref{fig:integrator-xy4}, $R^{ZZ}(t)$ for both colors are identical and displacement symmetric for simultaneous XY4. Staggering introduces a phase offset that generates orthogonality between the two displacement symmetric $R^{ZZ}(t)$ while leaving both independently displacement symmetric. Orthogonality ultimately affords the cancellation of this error matrix component.

The third component we highlight is $(\alpha,\beta)=(X,Z)$, which captures the mixing between pulse errors generated on qubit $R$ and $Z$-preserving errors from qubit $B$. Here, we have a product of $R^{ZX}(t)$, a displacement antisymmetric function, and $R^{ZZ}(t)$, a displacement symmetric function. The product remains displacement antisymmetric, and therefore integrates to zero for both simultaneous and crosstalk-robust sequence types. This reasoning extends to $(\alpha,\beta) = (Z,X), (Y,Z), (Z,Y)$ as well.

The behavior discussed above extends beyond $X$ and $Y$ pulses. We convey this characteristic here using a CR version of KDD, in which two KDD sequences are interleaved at the composite pulse layer. More specifically, consider two composite pulses $P_A$ and $P_B$ whose constituent pulses are described by ordered sets of phase angles $\Phi_A$ and $\Phi_B$, respectively. Assuming that the sizes of the sets are equal, we define the staggering operator $\mathcal{S}$ for composite pulses as
\begin{equation}
    \mathcal{S}(P_A,P_B) =\left\{
    \begin{array}{ccc}
    \prod_{\phi_j \in \Phi_A} &(\pi)_{\phi_j} &f_p\\
    \prod_{\phi_j \in \Phi_B} &f_p &(\pi)_{\phi_j} 
    \end{array}
    \right..
\end{equation}
In Fig.~\ref{fig:integrator-kdd}, this notation is utilized at the top of the column to further illustrate the sequence.

In investigating the control matrix elements for CR-KDD, we observe similar behavior to that of CR-XY4. Namely, the control matrix elements purely due to pulse errors remain orthogonal despite the increased complexity of the sequence. This is solely due to the fact that qubit $R$ and qubit $B$ are never subject to simultaneous pulsing. Similarly, $(\alpha,\beta)=(Z,Z)$ continues to exhibit displacement symmetric characteristics with a phase offset providing the necessary orthogonality for cancellation. Lastly, we note that the cross-terms such as $(\alpha,\beta)=(X,Z)$ again display the necessary displacement antisymmetry to achieve cancellation of the error matrix element in both the CR and non-CR implementations. The results of this analysis are summarized in Fig.~\ref{fig:integrator-kdd}.

\subsection{Heterogeneous Sequence Composition}
\begin{figure*}
    \centering
    \includegraphics[width=\textwidth]{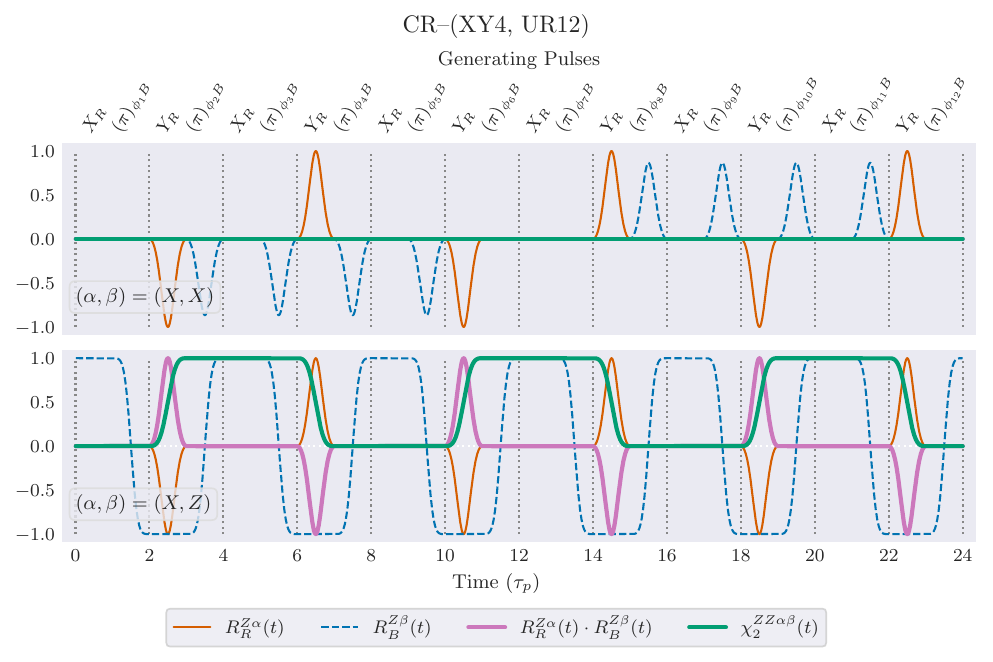}
    \caption{Selected pairs of elements from the control matrices of a heterogeneous sequence: CR-XY4 (red-orange) and CR-UR12 (blue). Results are shown for bounded control (DRAG pulses) through one cycle, their product, and the 2-local error matrix, $\chi_2^{ZZ\alpha\beta}(t)$. Orthogonality and symmetry continue to cancel errors despite the distinct sequence composition.}
    \label{fig:integrator-xyur}
\end{figure*}

Here, we demonstrate that the CR protocol remains valid even when composing distinct sequences. We display this feature using a composition of XY4 with UR12, the UR sequence with $\Phi_{\text{UR12}}=(0, 1, 3, 0, 4 , 3, 3, 4, 0, 3, 1, 0)\frac{\pi}{3}$. Three repetitions of XY4 are used to match the twelve pulses required to implement UR12. In Fig.~\ref{fig:integrator-xyur}, we highlight a subset of control matrix and error matrix elements for this heterogeneous sequence that we refer to as CR-(XY4,UR12).

Control matrix elements share many features with previously discussed sequences. For example, the elimination of  $(\alpha,\beta)=(X,X), (X,Y), (Y,X), (Y,Y), (Z,Z)$ follows directly from pulse staggering. As such, we will focus our discussion on the cross-terms, where $(\alpha,\beta)=(X,Z), (Y,Z), (Z,X), (Z,Y)$.

The repetition of XY4 results in many commonalities between CR-XY4 and CR-(XY4,UR12) control matrix elements. For example, consider $R^{ZZ}(t)$ for the UR12 sequence. It exhibits the same toggling behavior with each pulse as observed for XY4. Furthermore, it maintains the same displacement symmetry. As such, when combined with XY4 in the staggering protocol, the dynamics of $\chi_2^{ZZXZ}(t)$ appear nearly identical to the error matrix of CR-XY4. Note the three repetitions, however, due to increased cycle time. The resulting effect is the same: a cancellation of the error matrix element at the end of sequence.

For $(\alpha,\beta)=(Z,X), (Z,Y)$, we investigate the symmetries in $R_B^{Z\beta}(t)$, $\beta\in\{X,Y\}$ for UR12 applied to qubit $B$. Without any delay between pulses, UR12 would generate these control matrix elements as mirror antisymmetric. However, staggering pads delays between the pulses from one side. When UR12 is staggered, the control matrices $R_B^{ZX}(t)$ and $R_B^{ZY}(t)$ become displacement antisymmetric.  Therefore, under staggering XY4 with UR12, we are left with displacement symmetric $R_R^{ZZ}(t)$ and the displacement antisymmetric $R_B^{Z\beta}(t)$, which generates orthogonality.

The examples above substantiate our claim that $ZZ$ crosstalk robustness via pulse staggering extends beyond XY4. The source of the robustness can be attributed to interplay between the control matrix elements. In some cases, the matrix elements maintain orthogonality due to a local cancellation of the product of control matrix elements at each time $t$. In other cases, it is symmetry generated by the staggering that affords orthogonality. The latter observation relies in part on $R^{ZZ}(t)$ toggling between $\pm 1$ while maintaining displacement symmetry. This toggling is a key feature of utilizing sequences of $\pi$-pulse, and it is likely one of the primary sources for noise suppression. Extension of this protocol beyond $\pi$-pulses likely necessitates further consideration regarding how one can maintain symmetries while executing more generic pulses.

\section{\label{sec:experiments}Demonstrations on quantum processors}
We explore the efficacy of the staggered protocol further via demonstrations on quantum hardware. In particular, we perform a comparison between simultaneous and CR variants of sequences discussed in Sec.~\ref{sec:dd-detail}. We utilize superconducting qubit devices provided by the IBMQP and perform demonstrations on up to 20 qubits. We show that pulse staggering significantly outperforms simultaneously pulse implementation for all DD protocols. 

\subsection{\label{ssec:methods}Methods}
Our main demonstrations focus on preserving the state of collections of $n$ qubits. The methods tested are idling (IDLE), where the qubits are allowed to evolve freely according to their internal dynamics, simultaneously-driven DD (SIM-DD), and crosstalk-robust DD (CR-DD). 

Each DD method is tested over increasing \emph{protection durations} of wall time $t \in [0, T]$, where we take wall time to mean the elapsed time in microseconds. To standardize the number of pulses applied across the different DD methods, for a DD sequence generated by $\Phi_{\text{DD}}$, we take SIM-DD to be generated from Eq.~\eqref{eq:sim-dd} and CR-DD to be generated from Eq.~\eqref{eq:cr-dd},
both having sequence duration $\tau_c=2|\Phi_\text{DD}|\tau_p$. Once again, in this naming convention, the generating DD method's name is substituted for ``DD" to form the SIM or CR variant of that method (e.g., SIM-XY4 and CR-UR10).

\subsubsection{State Preparation Specifications}
Each circuit consists of three subcircuits: state encode, suppression, and state decode. The qubits are initialized in the ground state $\ket{0}^{\otimes n}$. State encoding involves applying a unitary operation $U_{\text{enc}}=\substack{ \bigotimes\\v \in V} U_{\ket{\psi_v}}$, which simultaneously prepares the $n$ qubits in the state $\ket{\psi_{\text{enc}}} = \substack{ \bigotimes\\v \in V} \ket{\psi_v}$, where the $\ket{\psi_v}$ need not be the same. The qubits are then subject to a suppression protocol or idle evolution before decoded using $U_{\text{dec}}=U_{\text{enc}}^\dagger$. By subsequently measuring in the computational basis we determine how well the prepared state was protected over the protection duration. Ideally, the output state of the experiment is the ground state $\ket{0}^{\otimes n}$.

The DD protocols are evaluated on sets of embeddings of four system sizes, $n=5, 10, 15,$ and $20$ qubits. Each embedding is placed as an $n$-qubit path graph in the IBM device topologies. We intentionally avoid embeddings that include edges in a device's topology with a CNOT gate error of 100$\%$ at compile time. We do this in case a non-functional gate is indicative of a connection that does not mediate crosstalk in the same way as the other connections.

\begin{figure*}[t]
    \centering
    \includegraphics[width=0.95\linewidth]{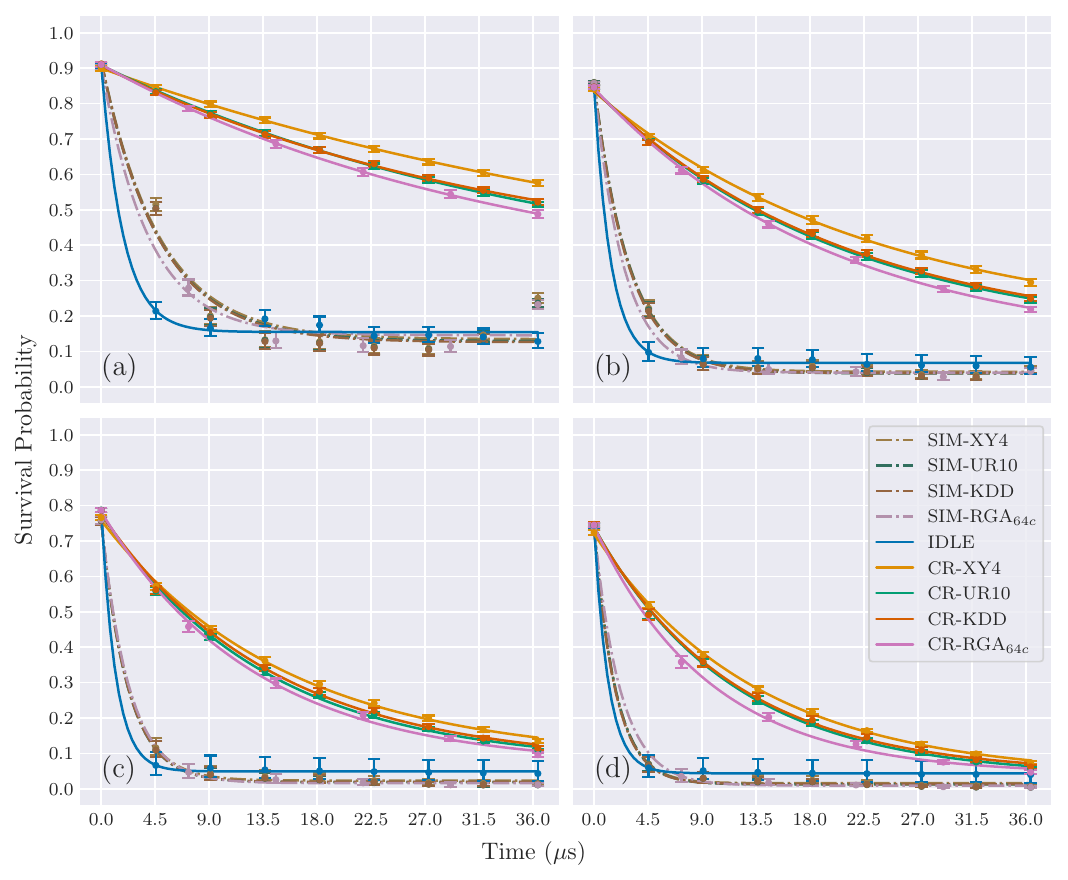}
    \caption{Performance comparisons of SIM-DD versus CR-DD implemented on  \sherbrooke{}. Mean survival probabilities are computed over both initial states and embeddings with 95\% confidence intervals of the mean. Overlaid curves are fit to the function of Eq.~\eqref{eq:expfit}. Subplots show data corresponding to different embeddings: (a) 19 embeddings of system size $n=5$, (b) 13 embeddings of $n=10$, (c) 8 embeddings of $n=15$, and (d) 7 embeddings of $n=20$. The results indicate that CR-DD consistently outperforms SIM-DD regardless the underlying DD protocol.}
    \label{fig:sherbrooke_fits}
\end{figure*}

For each $n$, we prepare two types of states. Type~1 states denote the case where all $n$ qubits are prepared in the same state. We define these states based on the six poles of the single-qubit Block sphere; thus, $\ket{\psi_\text{enc}}\in \{\ket{\pm z}^{\otimes n}, \ket{\pm x}^{\otimes n}, \ket{\pm y}^{\otimes n}\}$. Type 2 states correspond to non-uniform state preparations in which $\ket{\psi_v}$ are taken from an independent uniformly-random sampling of the six poles $\{\ket{\pm z}, \ket{\pm x}, \ket{\pm y}\}$ for each qubit. In our demonstrations, we consider 6 Type 1 states and 14 Type 2 states, resulting in a total of 20 distinct separable states. Previous studies have utilized similar constructions on single-qubit DD assessments and comparisons against logical encoding protocols as well~\cite{ezzell2023dynamicaldecoupling, quiroz2024dfs}.

\begin{figure*}[t]
    \centering
    \includegraphics[width=0.95\linewidth]{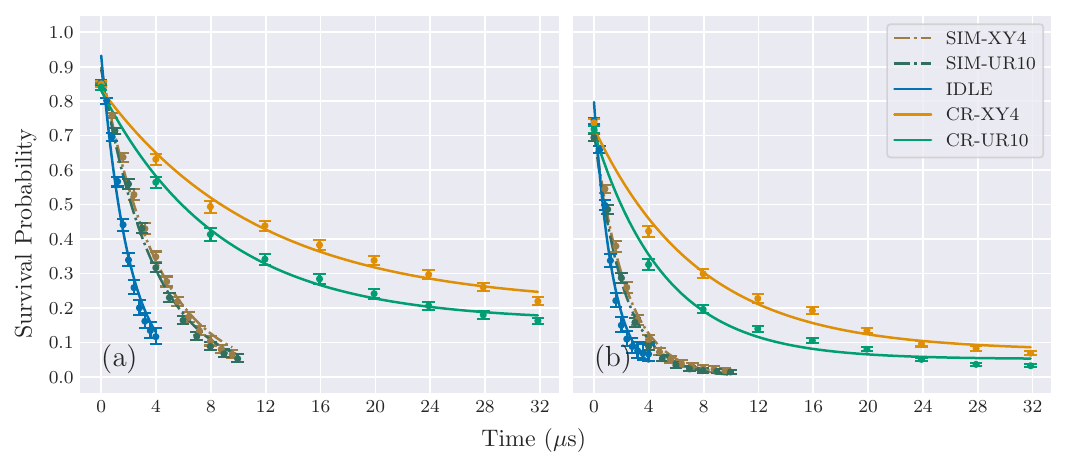}
    \caption{Performance comparisons of SIM-DD versus CR-DD implemented on \kyiv{}. Mean survival probabilities computed over both initial states and embeddings with error bars indicating the upper and lower ends of the 95\% confidence interval of the mean for each sample. Subplots show data taken with (a) 20 embeddings of system size $n=10$ and (b) 16 embeddings of $n=20$. Overlaid curves are fit to the function Eq.~\eqref{eq:expfit}. Data is taken at different intervals for different methods to better elucidate details regarding the decay in survival probability for IDLE and the simultaneous DD methods which occurs at shorter time than the CR methods. Once again, we observe CR-DD outperforming SIM-DD approaches.}
    \label{fig:kyiv_fits}
\end{figure*}

\subsubsection{Survival Probability}
We are interested in the probability of faithfully recovering the initial state with no error after some protection duration. To this end, we repeat the demonstration over a range of protection durations. In IDLE demonstrations, this corresponds to allowing the qubits to evolve freely for increasing periods of time. In the case of DD, this implies increasing the number of sequence repetitions. For a given protection duration, a circuit is repeated 1000 times to collect measurement statistics. The survival probability---the probability of returning to the ground state at the end of the circuit---is given by 
\begin{equation}
    P_0(T)=\braket{0\cdots 0| \rho_{\text{out}}(T)|0\cdots 0},
    \label{eq:survival-prob}
\end{equation}
where $\rho_{\text{out}}(T)$ is the output state of the circuit for a given protection duration $T$. Given that the ideal output state is equivalent to all qubits returning to the ground state, we estimate the survival probability as the probability of obtaining the output state $\ket{0\cdots 0}\bra{0\cdots 0}$ in the number of shots taken. Note that this metric rejects samples with a 1 on any qubit, a strict requirement as the number of qubits increases. We refer to the survival probabilities resulting from the set of protection durations for a given initial state and qubit embedding as a trace.

A focus of our comparison will be the characteristic time of the decrease of survival probability, which we call $\tau_\gamma$. In the subsequent subsections, we will extract an estimate of the probability decay rate by fitting the function
\begin{equation}\label{eq:expfit}
f(t)=Ae^{-\gamma t} + c
\end{equation}
to the survival probability traces. The parameter $A\in[0,1]$ allows the fitting procedure to scale the exponential decay to the possibly non-zero asymptotic value $c\in[0,1]$, while $\gamma$ denotes the decay rate in units of $\mu$s$^{-1}$. We obtain the characteristic time $\tau_\gamma = 1/\gamma$, the time at which the mean survival probability has been reduced from its starting point by $A/e$, or a factor of $1/e$ within its range. Below, we will compare the characteristic times across SIM and CR variants of DD protocols, embedding sizes, and quantum devices.

\subsection{\label{subsec:devices}Devices}
Demonstrations were carried out on three IBM devices. Secs.~\ref{subsec:state-prep-avg}, \ref{subsec:qubit-embedding}, and \ref{subsec:idle-padding} utilize \sherbrooke{} and \kyiv{}, two IBM devices based on the Eagle r3 architecture. Both devices are composed of 127 fixed-frequency transmons with fixed couplers~\cite{chow2011ffqubit, chow2012ffqubits}. It is known that such devices suffer from parasitic $ZZ$ crosstalk~\cite{krantz2019quantumengineers, tripathi2022suppressioncrosstalk,zhou2023crctrl, seif2024suppressingcorrelated,fors2024comprehensiveexplanation} and thus instantiate a relevant architecture for investigating the efficacy of our protocol. We utilize demonstrations for both devices to investigate variability in the performance of the suppression methods.

In Sec.~\ref{subsec:tunable}, we switch focus to \marrakesh{}, a Heron r2 architecture device based on fixed-frequency transmons with tunable coupling~\cite{stehlik2021tunable}. As a subsequent generation to the Eagle architecture, Heron devices are designed to suppress parasitic $ZZ$ errors. We utilize our DD protocol to study the intrinsic suppression abilities of this architecture. Further information regarding device properties can be found in Appendix~\ref{sec:devices}.

To present the large amount of data collected, we have prepared visualizations and tables in terms of summary statistics. Often, a figure (table) will contain data for a round of testing on one device, and data for different qubit embedding sizes $n$ are presented in subplots (table sections). Comparing performance across embedding sizes on the same device provides insight into the trend of a suppression technique's effectiveness as the number of qubits increases.

\begin{figure*}[t]
    \centering
    \includegraphics[width=0.95\linewidth]{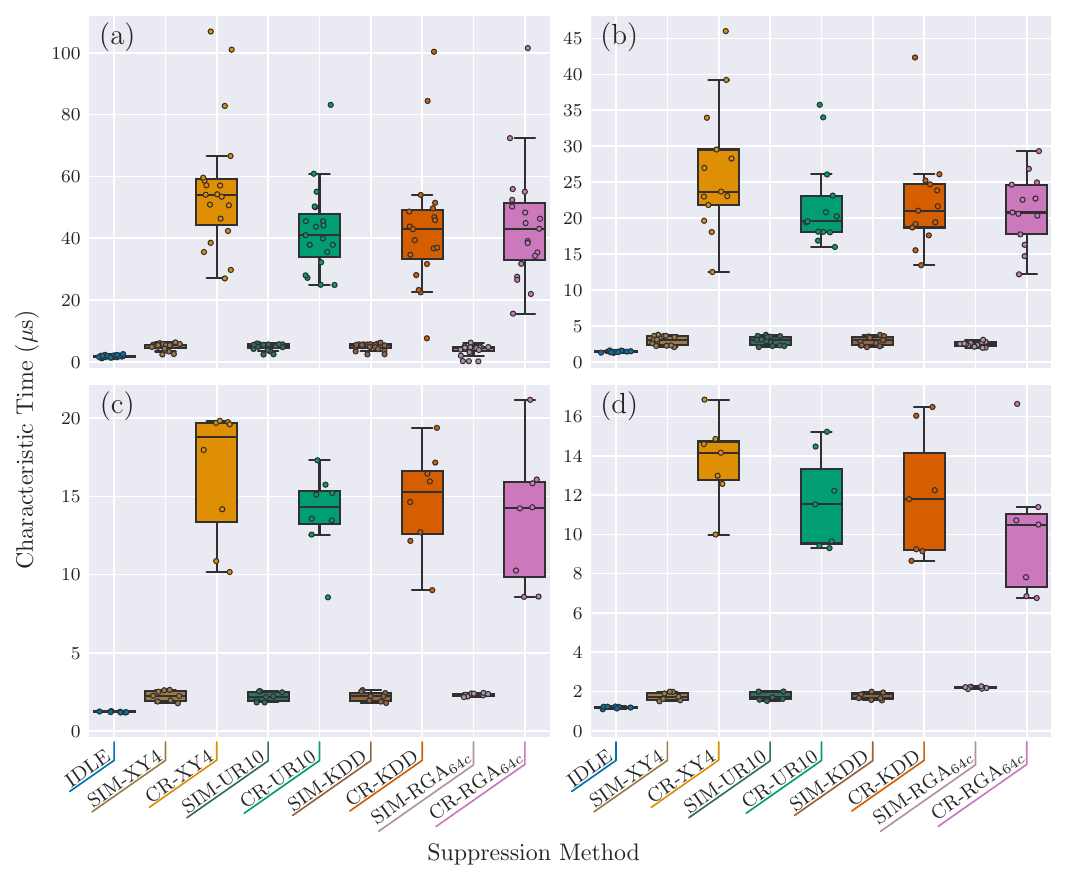}
    \caption{Comparison of characteristic times of survival probability for IDLE, SIM-DD, and CR-DD on \sherbrooke{}.  Subplots show data for different embeddings: (a) 19 embeddings of system size $n=5$, (b) 13 embeddings of $n=10$, (c) 8 embeddings of $n=15$, and (d) 7 embeddings of $n=20$. Each data point shows the characteristic time of the average survival probability taken over initial states for each embedding. Boxes show the median, 25th and 75th percentiles, and whiskers are drawn up to {1.5$\times$IQR}. Overall, we observe significant increases in the characteristic time for CR-DD for a majority of the embeddings considered.}
    \label{fig:sherbrooke_times}
\end{figure*}

\subsection{Multi-qubit State Preservation: Average Device Performance}
\label{subsec:state-prep-avg}
Here, we compare the performance the SIM and CR protection methods on \sherbrooke{} and \kyiv{} using the DD protocols overviewed in Sec.~\ref{sec:dd-detail}. First, we focus on the survival probability averaged over all embeddings and initial states for a given $n$. In Fig.~\ref{fig:sherbrooke_fits} we show results for this comparison for \sherbrooke{} as a function of cycle time.  Fig. \ref{fig:kyiv_fits} shows results for a comparison of a particular set of sequences: IDLE, XY4, and UR10 (both SIM and CR variants). Data points denote the mean survival probability, while the error bars show 95\% confidence intervals obtained from a bootstrap resampling (10,000 samples with replacement) of the survival probabilities. Each subplot includes curves obtained from fitting Eq.~\eqref{eq:expfit} to the data obtained for IDLE, SIM-DD, and CR-DD. In addition to the quantitative utility of curve fitting for computing characteristic times, the fits serve as indicators of the difference in performance between the suppression methods.

\begin{figure*}[t]
    \centering
    \includegraphics[width=0.95\linewidth]{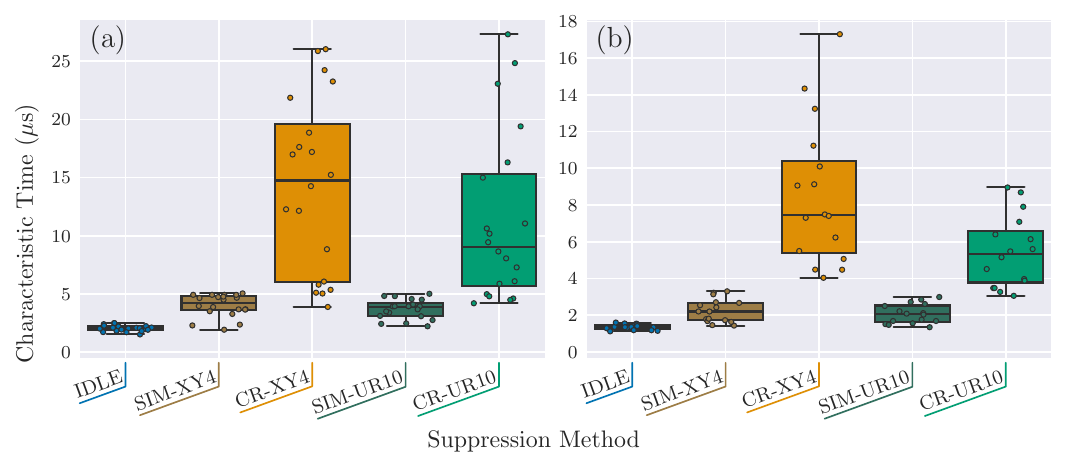}
    \caption{Comparison of characteristic times of survival probability for IDLE, SIM-DD, and CR-DD on \kyiv{}. Subplots show data taken with different embeddings: (a) 20 embeddings of system size $n=10$ and (b) 16 embeddings of $n=20$. Each data point shows the characteristic time of the average survival probability taken over initial states for each embedding. Boxes show the median, 25th and 75th percentiles, and whiskers are drawn with length up to {1.5$\times$IQR}. While CR-DD typically outperforms SIM-DD, this device exhibits greater variability in performance with respect to embedding.}
    \label{fig:kyiv_times}
\end{figure*}

In Fig.~\ref{fig:sherbrooke_fits}, panels (a)-(d) correspond to $n=5,10,15,20$ qubit comparisons, respectively. In order to effectively assess the performance of each DD protocol over a wide range of qubits on \sherbrooke{}, we consider multiple embeddings for each $n$. For $n=5$, we utilize 19 embeddings, while for $n=10, 15, 20$, we use 13, 8, and 7, respectively. To ensure equal resources between approaches, each data point corresponds to an equivalent number of pulses. As such, data is collected at intervals of a common multiple of the DD cycle durations. The pulse duration on \sherbrooke{} is {$\tau_p=56.\bar8$ns}, which yields a wall time of {4.55$\mu$s} between data points for IDLE, XY4, UR10, and KDD (40 pulses and delays) and of {7.28$\mu$s} between data points for RGA$_{64c}$ (64 pulses and delays). The final data points for all methods are aligned at a wall time of 640 pulse durations, when all of the active methods (i.e., those other than IDLE) have applied 320 pulses of DD to each qubit.

Demonstrations performed on \sherbrooke{} convey an advantage from pulse staggering, indicating suppression of $ZZ$ errors for all CR protocols. We do not observe a major qualitative difference between the traces of the CR methods or between the traces of the SIM methods, suggesting that the choice of the generating DD method is less important on this device than making the DD method robust to spatially-correlated noise. However, CR-XY4 does reliably have the highest (best) $\tau_\gamma$ of the CR methods. We find that, for $n=5$, CR-XY4 obtains {$\tau_\gamma=54.1\mu$s}, compared to {$\tau_\gamma=5.32\mu$s} for SIM-XY4; a $10.2\times$ improvement. Similarly, we determine CR-XY4 for $n=10,15,20$ as {$\tau_\gamma=23.7\mu$s}, {$18.8\mu$s}, and {$14.2\mu$s} which constitute $7.68\times$, $8.38\times$, and $8.15\times$ improvements over SIM-XY4, respectively.

\begin{table*}[t]
    \centering
    \begin{tabular*}{5.5in}{@{\extracolsep{\stretch{1}}}*{7}{c}@{}}
        \hline\hline
        $n$ & Method & SIM Median $\tau_\gamma$ (IQR) ($\mu$s) & CR Median $\tau_\gamma$ (IQR) ($\mu$s) & SIM / IDLE & CR / SIM \\
\hline
5 & IDLE & 1.88 (0.515) & – & – & – \\
& XY4 & 5.32 (1.03) & 54.1 (14.8) & 2.83 & 10.2 \\
& UR10 & 5.15 (1.15) & 41.0 (13.9) & 2.74 & 7.95 \\
& KDD & 5.29 (1.2) & 42.9 (16.0) & 2.81 & 8.11 \\
& RGA$_{64c}$ & 4.41 (1.37) & 43.1 (18.3) & 2.35 & 9.76 \\
\hline
10 & IDLE & 1.4 (0.116) & – & – & – \\
& XY4 & 3.09 (1.31) & 23.7 (7.73) & 2.2 & 7.68 \\
& UR10 & 3.05 (1.21) & 19.5 (5.06) & 2.17 & 6.41 \\
& KDD & 2.99 (1.1) & 21.0 (6.01) & 2.14 & 7.02 \\
& RGA$_{64c}$ & 2.49 (0.461) & 20.8 (6.93) & 1.78 & 8.33 \\
\hline
15 & IDLE & 1.21 (0.0512) & – & – & – \\
& XY4 & 2.24 (0.591) & 18.8 (6.37) & 1.85 & 8.38 \\
& UR10 & 2.19 (0.549) & 14.3 (2.11) & 1.8 & 6.56 \\
& KDD & 2.21 (0.521) & 15.3 (4.06) & 1.82 & 6.92 \\
& RGA$_{64c}$ & 2.33 (0.131) & 14.3 (6.06) & 1.92 & 6.12 \\
\hline
20 & IDLE & 1.22 (0.0639) & – & – & – \\
& XY4 & 1.74 (0.351) & 14.2 (1.95) & 1.43 & 8.15 \\
& UR10 & 1.71 (0.384) & 11.5 (3.81) & 1.4 & 6.76 \\
& KDD & 1.67 (0.281) & 11.8 (4.95) & 1.37 & 7.06 \\
& RGA$_{64c}$ & 2.23 (0.0713) & 10.5 (3.71) & 1.83 & 4.72 \\
    \hline\hline\end{tabular*}
    \centering
    \caption{Comparison of median characteristic times for each DD method implemented on \sherbrooke{}. Median characteristic time is taken over all embeddings for each system size $n$. The characteristic time for each embedding is estimated via a fit to the survival probability averaged over initial states Eq.~\eqref{eq:expfit}. The data shown here complements the data displayed in Fig.~\ref{fig:sherbrooke_times}. Results indicate a substantial improvement from CR-DD relative to both IDLE and SIM-DD.}
    \label{tab:sherbrooke_times}
\end{table*}

\begin{table*}[t]
    \centering
    \begin{tabular*}{5.5in}{@{\extracolsep{\stretch{1}}}*{7}{c}@{}}
        \hline\hline
        $n$ & Method & SIM Median $\tau_\gamma$ (IQR) ($\mu$s) & CR Median $\tau_\gamma$ (IQR) ($\mu$s) & SIM / IDLE & CR / SIM \\
\hline
10 & IDLE & 2.03 (0.311) & – & – & – \\
& XY4 & 4.22 (1.19) & 14.7 (13.6) & 2.08 & 3.5 \\
& UR10 & 3.89 (1.09) & 9.04 (9.68) & 1.92 & 2.32 \\
\hline
20 & IDLE & 1.35 (0.182) & – & – & – \\
& XY4 & 2.2 (0.963) & 7.45 (5.0) & 1.63 & 3.38 \\
& UR10 & 2.06 (0.876) & 5.31 (2.79) & 1.52 & 2.58 \\
        \hline\hline\end{tabular*}
    \centering
    \caption{Comparison of median characteristic times $\tau_\gamma$ of survival probability of each suppression method on \kyiv{}. The median is taken over embeddings on \kyiv{}. The $\tau_\gamma$ for each embedding is estimated via a fit to the survival probability averaged over initial states Eq.~\eqref{eq:expfit}. The data shown here complements the data displayed in Fig.~\ref{fig:kyiv_times}. Results indicate a substantial improvement from CR-DD relative to both IDLE and SIM-DD.}
    \label{tab:kyiv_times}
\end{table*}

A similar comparison is shown for \kyiv{} in Fig.~\ref{fig:kyiv_fits}, including IDLE and both SIM and CR variants of XY4 and UR10. To get a better picture of the decay profiles for IDLE and SIM-DD, we collect data at shorter wall times for these methods in this demonstration. Additionally, the pulse duration on \kyiv{} is {$\tau_p=49.\bar7$ns} which is shorter than on \sherbrooke{}. This timing yields a wall time of {$3.98\mu$s} ($80\tau_p$) between data points for IDLE, XY4, and UR10.

The trends in Fig.~\ref{fig:kyiv_fits} are similar to those in Fig.~\ref{fig:sherbrooke_fits} with some key differences. Similar to \sherbrooke{}, this data from \kyiv{} shows any DD method is on average better than IDLE. We also continue to see enhanced $\tau_\gamma$ for the CR-DD methods over the SIM-DD methods. However, there remains a distinction between the performances of CR protocols. CR-XY4 continues to dominate, with median $\tau_\gamma$ of {$\tau_\gamma=14.75\,\mu$s} ($n=10$) and {$\tau_\gamma=7.45\,\mu$s} ($n=20$). CR-UR10 on \kyiv{} has median $\tau_\gamma$ of {$\tau_\gamma=9.04\,\mu$s} ($n=10$) and {$\tau_\gamma=5.31\,\mu$s} ($n=20$), consistently worse performance than CR-XY4. 

We attribute the difference in performance between the DD methods to residual 1-local errors. Pulse staggering suppresses the 2-local $ZZ$ errors in the first-order Magnus term completely for both CR-XY4 and CR-UR10. At the completion of the sequences, the error dynamics are dominated by any remaining 1-local error terms. These terms are given by non-zero error matrix elements described by Eq.~\eqref{eq:1-local-err-matr}. Upon numerical examination of the 1-local error matrix components, we find that CR-XY4 possesses remaining errors along $(\mu,\alpha)=(X,Z), (Y,Z)$. In contrast, CR-UR10 yields non-zero error contributions for $(\mu,\alpha)=(X,X), (X,Y), (Y,X), (Y,Y)$. Both indicate errors due to longitudinal coupling from the $B^X_v$ and $B^Y_v$ environment operators. The distinction lies in the number of contributing terms and overall magnitudes for the error matrix elements. For an equal number of pulses, the dominant 1-local error matrix elements for CR-UR10 are approximately {$1.85\times$} larger than those of CR-XY4. We conjecture that this difference in residual error accounts for the distinction between the two DD methods and further indicates that \kyiv{} is more susceptible to longitudinal errors than \sherbrooke{}, consistent with the shorter $T_1$ times observed on \sherbrooke{}.

\subsection{Multi-Qubit State Preparation: Variability with Qubit Embedding}
\label{subsec:qubit-embedding}
As a further investigation, we examine the distribution of $\tau_\gamma$ with respect to embedding in Figs.~\ref{fig:sherbrooke_times}~and~\ref{fig:kyiv_times}. Each data point denotes a $\tau_\gamma$ obtained from a fit of Eq.~\eqref{eq:expfit} to the state-averaged survival probability for each embedding. A longer $\tau_\gamma$ is better as it indicates a slower decrease of the survival probability. The data points are accompanied by box and whisker plots, where a box denotes first and third quartiles and whiskers are drawn to the furthest data point within 1.5 times the inter-quartile range (IQR) and are otherwise omitted if no such data point exists. The solid line within the box corresponds to the median. Each panel contains data for embeddings with a different system size in correspondence to Figs.~\ref{fig:sherbrooke_fits}~and~\ref{fig:kyiv_fits}.

Fig.~\ref{fig:sherbrooke_times} reveals the relative advantages of DD for \sherbrooke{}. In aggregate, the SIM methods consistently outperform IDLE in embeddings across the device. However, there are one or two outliers for each of the SIM-DD methods that overlap with the range of $\tau_\gamma$ observed for IDLE on 5-qubit embeddings. We find another consistent increase in sequence performance when introducing pulse staggering for all embeddings, indicating that (1) crosstalk is prevalent across the device and (2) the enhancement from the CR protocol is not qubit dependent.

We show similar results for \kyiv{} in Fig.~\ref{fig:kyiv_times}. We still observe an improvement from DD over IDLE and CR-DD sequences over SIM-DD. There continues to be a small number of outlier embeddings with SIM-DD performance that falls within the IQR of IDLE, and a small number of outlier embeddings with CR-DD performance that falls within the IQRs of IDLE and/or SIM-DD. The overlap between distributions is reduced as system size is increased.

We tabulate the values and relative improvements of median $\tau_\gamma$ taken over initial-state-averaged survival probability of embeddings for each suppression method in Tables~\ref{tab:sherbrooke_times}~and~\ref{tab:kyiv_times}. Each table is divided into four sections corresponding to results of the 5-, 10-, 15-, and 20-qubit embedding circuits. In each section, we list the median $\tau_\gamma$ of the IDLE method, which all other methods are compared to. The IDLE test was repeated when both the SIM data was collected and again when the CR data was taken. The better performing IDLE $\tau_\gamma$ is always used in the comparisons. For each DD sequence, we list the median $\tau_\gamma$ of both its SIM and CR variants. The medians are the same as those from the box plots in Figs.~\ref{fig:sherbrooke_times}~and~\ref{fig:kyiv_times}. To compare characteristic times we compute the ratios of times rather than working with absolute differences between them. The ratios represent the relative improvements of each CR method over the corresponding SIM method and over the IDLE performance for each system size $n$.

In Table~\ref{tab:sherbrooke_times}, results for \sherbrooke{} are summarized. CR-XY4 consistently has the best enhancement in $\tau_\gamma$ compared with IDLE and SIM-XY4, with up to $28.7\times$ and $10.2\times$ improvements in $\tau_\gamma$, respectively. In one specific case, $n=10$, RGA$_{64c}$ obtains the highest enhancement for the CR variant over SIM; however, this is not a typical scenario. Comparing the ratio of CR to SIM decay rates (last column of Table~\ref{tab:sherbrooke_times}), we find that for larger embeddings ($n=15$ and $n=20$), RGA$_{64c}$ exhibits the smallest improvement relative to the other DD methods. Nevertheless, its improvement is still significant. Increases in the characteristic time by factors of $6.12\times$ ($n=15$) and $4.72\times$ ($n=20$) are observed.

XY4 continues to be the highest performer on \kyiv{} as well, as illustrated in Table~\ref{tab:kyiv_times}. The best improvement over IDLE is generally CR-XY4. The UR10 sequences performed similarly to the XY4 sequences, though with a consistently shorter $\tau_\gamma$. These results continue to support the improvement afforded by DD with pulse staggering over simultaneous pulses on these devices.

\begin{figure}[t]
    \centering
    \includegraphics[width=\linewidth]{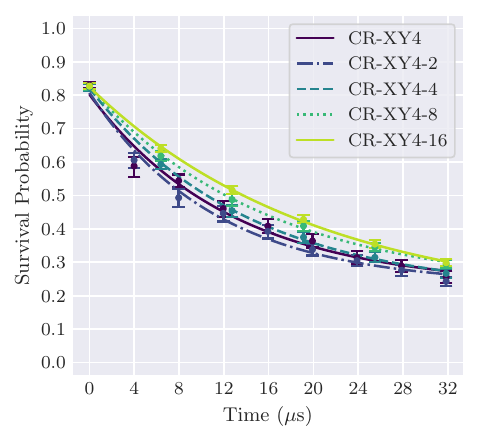}
    \caption{Comparison of mean survival probabilities for padded CR-XY4 with varied cycle wall times on \kyiv{}. Data is aggregated over results from testing ten embeddings of system size $n=10$ qubits with 20 initial states each. The suffix {-$k$} denotes CR-XY4 padded symmetrically with equal delays around the pulses according to Eq.~\eqref{eq:sym-pad-cr-dd}. DD performance is relatively consistent, with a preference towards fewer pulses and longer interpulse delays.}
    \label{fig:kyiv_padded}
\end{figure}

\subsection{Effects of Idle Padding}
\label{subsec:idle-padding}
In Sec.~\ref{sec:eng-suppression}, we remark that pulse staggering maintains crosstalk robustness even when inter-pulse delays are included. This feature is particularly relevant in scenarios where DD is used to protect idling qubits in quantum algorithms~\cite{pokharel2018demonstrationfidelity, kim2023scalableerror, pokharel2024betterthanclassicalgrover, singkanipa2024demonstrationalgorithmic}. The CR-DD extension builds upon previous work, enabling a new approach for protecting collections of idle qubits in algorithms when $ZZ$ errors are a dominant source of noise. Importantly, the ability to stagger various types of sequences adds further options that can be exploited by recently developed DD compiler pass methodologies~\cite{seif2024suppressingcorrelated, coote2024resourceefficientcontextaware}. Here, as a first step toward such integration, we illustrate the potential utility of padding additional delay between pulses via a demonstration on \kyiv{}.

\begin{figure*}[t]
    \centering
    \includegraphics[width=\linewidth]{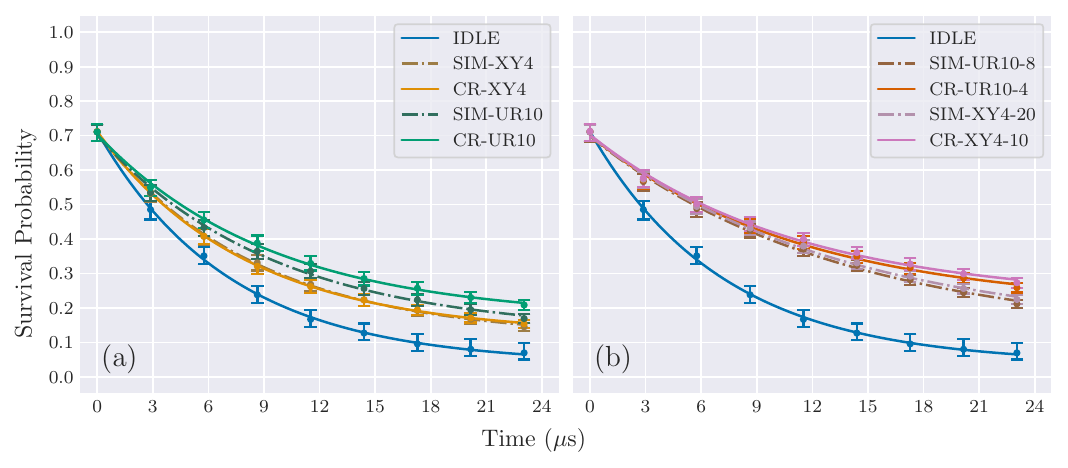}
    \caption{Performance comparisons of SIM-DD versus CR-DD implemented on \marrakesh{}. Data points denote mean survival probabilities computed over the 20 initial states (6 Type 1 and 14 Type 2 states) and ten embeddings of system size $n=10$. Error bars indicate the upper and lower ends of the 95\% confidence interval of the mean. Subplot (a) shows IDLE and unpadded SIM and CR variants of both XY4 and UR10. Subplot (b) shows IDLE, as well as UR10 and XY4 (SIM and CR variants) asymmetrically padded to $\tau_c=80\tau_p$ according to Eq.~\eqref{eq:asym-pad-cr-dd}. The same IDLE data appears on both plots as this data was taken in one batched job. Unlike the results for the fixed-coupler devices, we do not observe a significant benefit from CR-DD.}
    \label{fig:marrakesh_fits}
\end{figure*}

In Fig.~\ref{fig:kyiv_padded}, we display a demonstration of CR-XY4 with additional symmetric idle padding in accordance with Eq.~\eqref{eq:sym-pad-cr-dd}. Results are shown for ten embeddings of system size $n=10$ and with 20 initial states following the Type 1 and Type 2 state preparation. The survival probability is averaged over initial states and embeddings and plotted as a function of the wall time. As before, curves of the form Eq.~\eqref{eq:expfit} are fit to the mean trace. Data is shown for $k=1,2,4,8,16$, where the number of cycles between data points is varied such that all of the methods are executed to a total protection of 640 pulse durations.

Compared with the minimal CR-XY4 sequence, the resulting curves have slightly reduced ($k=2$) or improved ($k=4, 8, 16$) performance in terms of survival probability, indicating that the sequences with additional padding are still robust to $ZZ$ crosstalk. As the only difference between these sequences is the number of pulses per wall time, this result suggests that the there is a benefit to applying the CR protocol with padding. With it, we can continue to obtain a given survival probability over the same protection duration with fewer pulses (or alternatively, over a longer protection duration with the same number of pulses). This result is consistent with previous DD studies that have noted the utility of using fewer pulses due to accumulating pulse error~\cite{ezzell2023dynamicaldecoupling}.

\subsection{Indirect Verification of Crosstalk Suppression in Tunable-Coupler Devices}
\label{subsec:tunable}
In recent years, DD has been employed for characterization in addition to error suppression. By carefully designing a DD sequences, one can tailor the frequency response of the qubit to effectively learn spectral characteristics of the noise. This technique has become known as quantum noise spectroscopy~\cite{bylander2011noisespectroscopy, alvarez2011measuringspectrum, pazsilva2017qns}. CR-DD methods have previously been utilized to combat $ZZ$ errors and improve the efficacy of noise spectroscopy directly~\cite{zhou2023crctrl}. Here, we consider an alternative application for CR-DD that facilitates direct detection of $ZZ$ errors, or alternatively, verification of suppression. 

\begin{figure}[t]
    \centering
    \includegraphics[width=\linewidth]{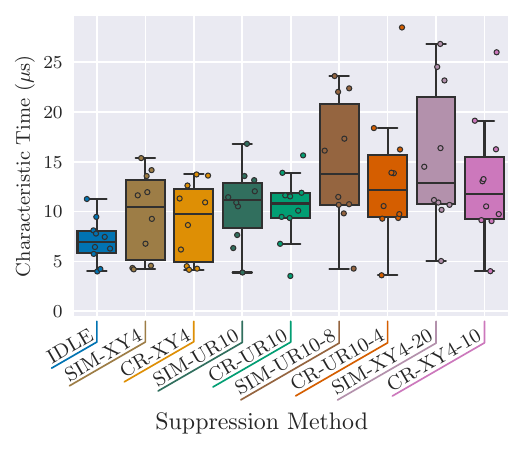}
    \caption{Comparison of characteristic times of survival probability for IDLE, SIM-DD, and CR-DD on \marrakesh{}. Data is taken over ten embeddings of system size of $n=10$ qubits. Each embedding is subject to 20 initial states (6 Type 1 and 14 Type 2 states). Complementing Fig.~\ref{fig:marrakesh_fits}, substantial benefits from CR-DD are not observed. SIM-DD typically performs equivalently, if not better than CR-DD protocols.}
    \label{fig:marrakesh_char_times}
\end{figure}

To this end, we apply CR-DD to qualitatively determine whether crosstalk is a major contributing source of noise in a device. Because the crosstalk robustness method can empower a DD sequence to cancel spatially correlated noise with low constant overhead, one can test both the SIM and CR versions of a DD sequence with identical single-qubit decoupling and observe whether there is a difference in the trends of survival probability between the methods.

\begin{table*}[t]
    \centering
    \begin{tabular*}{5.5in}{@{\extracolsep{\stretch{1}}}*{7}{c}@{}}
        \hline\hline
        $n$ & Method & SIM Median $\tau_\gamma$ (IQR) ($\mu$s) & CR Median $\tau_\gamma$ (IQR) ($\mu$s) & SIM / IDLE & CR / SIM \\
\hline
10 & IDLE & 6.95 (2.15)& – & – & – \\
& XY4 & 10.5 (8.05) & 9.79 (7.35) & 1.5 & 0.936 \\
& UR10 & 11.2 (4.51) & 10.8 (2.44) & 1.61 & 0.968 \\
& UR10-$k$ & 13.8 (10.2) & 12.2 (6.2) & 1.98 & 0.885 \\
& XY4-$k$ & 12.8 (10.7) & 11.8 (6.22) & 1.85 & 0.919 \\
        \hline\hline\end{tabular*}
    \centering
    \caption{Comparison of median characteristic times $\tau_\gamma$ of survival probability of each suppression method on \marrakesh{}. UR10-$k$ refers to SIM-UR10-8 and CR-UR10-4, and XY4-$k$ refers to SIM-XY4-20 and CR-XY4-10, which all have cycle duration $\tau_c=80\tau_\gamma$. The median is taken over ten embeddings of system size $n=10$ on \marrakesh{}. The $\tau_\gamma$ for each embedding is estimated as discussed in Table.~\ref{tab:sherbrooke_times}. The median characteristic times are quite similar among SIM-DD and CR-DD methods, suggesting that $ZZ$ errors are not a dominant error for this device.}
    \label{tab:marrakesh_times}
\end{table*}

If the CR methods tend to have higher survival probabilities and lower decay rates than the SIM methods, then it is reasonable to believe that there is spatially-correlated noise. However, if the SIM and CR methods have similar decay profiles, then the addition of crosstalk robustness did not have a meaningful effect. In this case, it is unlikely that spatially-correlated noise is a key contributor to decoherence in the device in question. We utilize this observation to investigate intrinsic $ZZ$ suppression afforded by the IBMQP's newest generation of devices based on tunable couplers. The advantage of using tunable couplers in superconducting qubit devices is that it should grant the ability to greatly diminish unwanted $ZZ$ interactions between fixed-frequency transmons~\cite{blais2003tunablecoupling}.

We apply the SIM and CR protocols to \marrakesh{}, a Heron r2 device equipped with tunable couplers. Our demonstration shown in Fig.~\ref{fig:marrakesh_fits} uses ten embeddings of system size $n=10$ qubits and the same 20 initial states (6 Type 1 and 14 Type 2 states) on which to perform state preservation. In panel (a), we compare XY4 and UR10 variants, while in panel (b), we consider both protocols with additional asymmetric idle padding according to Eq.~\eqref{eq:asym-pad-cr-dd}. The protocols are padded such that the same number of pulses is applied to each qubit for the SIM and CR variants over the same protection durations. The pulse duration on \marrakesh{} is {$\tau_p=36$ns}, a factor of $1.58\times$ and $1.38\times$ shorter than \sherbrooke{} and \kyiv{}, respectively.

We observe some qualitative similarities and differences to the results found for \sherbrooke{} and \kyiv{}. Analogous to the \sherbrooke{} and \kyiv{} results, the DD methods continue to increase the average survival probability over IDLE on \marrakesh{}. For the base CR-UR10 and both of the padded CR-XY4-10 and CR-UR10-4 methods, we continue to see an improvement over the SIM methods in the observed average survival probability across the tested protection durations. A departure from the Eagle r3 results is that there is no longer a wide gap between the mean survival probabilities of the SIM and CR methods. While the mean survival probability traces are higher for the three aforementioned CR methods over the SIM variants, the average survival probabilities of the XY4 variants were closely matched. We take this as an indication that spatially-correlated noise is being greatly reduced at the hardware level in this tunable coupler device.

The distributions of $\tau_\gamma$ for the methods over the ten embeddings is shown in Fig.~\ref{fig:marrakesh_char_times} and numerical data tabulated in Table~\ref{tab:marrakesh_times}. Although the survival probabilities are overall slightly higher for the CR methods, the observed $\tau_\gamma$ show there is about a 3\% to 11\% penalty in characteristic time when using CR-DD. This is in contrast to the strong advantage of the CR methods over the SIM methods demonstrated on the Eagle r3 devices. The best median $\tau_\gamma$ observed on \marrakesh{} is {SIM-UR10-8}, with  {$\tau_\gamma=13.8\mu$s.} However, when compared with the median $\tau_\gamma$ for $n=10$ on \sherbrooke{}, {$\tau_\gamma=23.7\mu$s}, and \kyiv{},  {$\tau_\gamma=14.7\mu$s}, we find that the median characteristic time of survival probability on \marrakesh{}  is $1.72\times$ shorter than on \sherbrooke{} and $1.07\times$ shorter than on \kyiv{}.

While the characteristic time captures the rate of decay in fidelity, it does not provide information about the magnitude of the survival probability. It is from the perspective of the latter that we observe a notable finding. Specifically, IDLE on \marrakesh{} outperforms both IDLE and SIM-DD on \sherbrooke{} and \kyiv{}. Despite these improvements afforded by the tunable-coupler device, the fixed-coupler are able to maintain higher survival probability magnitudes when utilizing CR-DD.

\begin{figure}[t]
    \centering
    \includegraphics[width=\linewidth]{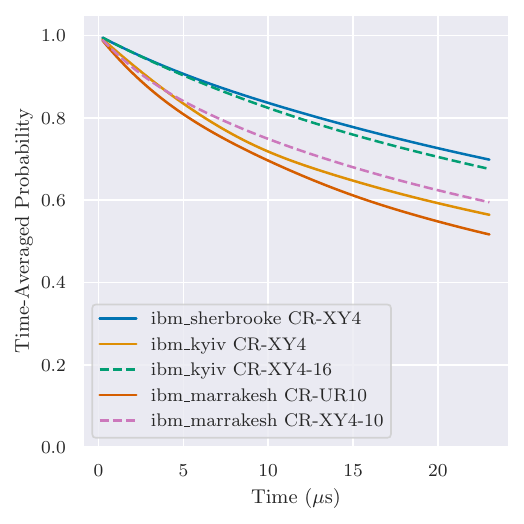}
    \caption{Comparison of cubic splines fit to SPAM-normalized mean traces of the best-performing methods on each device by final survival probability. The best CR methods on fixed-coupler devices outperformed any DD variant tested on the tunable-coupler device.}
    \label{fig:time_avg_probability}
\end{figure}

The separation between the best methods demonstrated on each device is made apparent by extending this analysis to the time-averaged survival probability. We obtain a smooth probability function $p_0(t)$ by fitting a cubic spline to the mean trace of each best-performing method for system size $n=10$ on each device. Using this continuous approximation of survival probability, the time-averaged survival probability is then
\begin{equation}
    \mathcal{P}_0(T) = \frac{1}{T} \int_0^T \frac{p_0(t)}{p_0(0)}\,dt,
\end{equation}
where we have normalized by the initial probability to account for state preparation and measurement (SPAM) error. These are presented as curves over {$T\in(0,23.04]\mu$s} in Fig.~\ref{fig:time_avg_probability}, showing CR-XY4 on \sherbrooke{} and \kyiv{}, CR-UR10 on \marrakesh{}, CR-XY4-16 on \kyiv{} and CR-XY4-10 on \marrakesh{}. The curves from the best methods on the fixed-coupling devices dominate those of \marrakesh{}.
Consequently, our demonstrations suggest that, under current technology specifications, it may be more favorable to leverage fixed-coupler architectures in conjunction with crosstalk-robust control (i.e., software-based) methods than to employ tunable-coupler architectures.

\section{\label{sec:conclusion}Conclusions}
In summary, we have further generalized a method for enabling robustness to spatially-correlated noise in DD. The approach relies on staggering DD pulses in time in order to suppress $ZZ$ errors and any effective pulse error induced by finite-duration pulses. The approach requires a total sequence cycle time only twice as long as the base sequence and does not demand additional cost in pulse number. Our approach applies to a large family of DD sequences and any two-colorable qubit topology.

Through numerical investigation, we have shown examples of how the method modifies DD sequences in order to satisfy first-order crosstalk robustness conditions. In particular, pulse staggering, specifically when applied to sequences of equally-spaced $\pi$-pulses, can satisfy particular symmetry conditions that enable cancellation of $ZZ$ errors and effective two-body interaction terms resulting from pulse error. The method is broadly applicable, even permitting for the composition of distinct DD sequences.

We complement our numerical study by demonstrations on the IBMQP. We draw on multiple high-performing single-qubit DD sequences and promote them to their crosstalk-robust variants. They are subsequently employed in state preservation experiments with up to 20 qubits, and across multiple devices and qubit embeddings. Demonstrations on fixed-coupler transmon architectures convey substantial improvements from our approach for all sequences studied.

In addition, the crosstalk-robust sequences are used as a detection method to verify hardware-enabled crosstalk suppression. Crosstalk-robust variants are applied to an IBMQP device with tunable couplers and shown to offer little advantage over their non-robust counterparts. This study provides evidence that crosstalk is indeed substantially mitigated at the hardware layer for this generation of devices. To the best of our knowledge, this is the first independent verification of this feature. 

Our analysis of DD on tunable-coupler devices further reveals that despite the existence of $ZZ$ crosstalk, fixed-coupler devices may achieve higher DD-protected survival probabilities than those of tunable-coupler devices. We observe this specifically when comparing CR-DD sequences across all three devices used in this study. This finding suggest that fixed-coupler architectures may remain a viable technology when supported by software infrastructure designed to address quantum crosstalk.

There are several directions we identify for future work. First, whether it is possible to extend the method beyond two-colorable topologies and retain robustness to spatially-correlated noise with bounded controls. Second, identification of sequences that eliminate spatially-correlated noise more generically, not just {$ZZ$,} while maintaining the flexible scheduling properties of short DD sequences with arbitrary extra padding. Third, while we have demonstrated the potential improvements of sequences resulting from our method on state preservation (i.e., memory tasks), an interesting avenue would be to study algorithmic tasks. Building a system with context-aware compilation~\cite{seif2024suppressingcorrelated, coote2024resourceefficientcontextaware} to embed DD sequences from our method into algorithms designed for bipartite topologies could enable exploration of a greater diversity of sequences and lead to more performant algorithms.

\section{Acknowledgments}
We thank Yasuo Oda for his insightful comments and Jacob Young for fruitful discussion. This work was supported in part by the U.S. Department of Energy, Office of Science, Office of Advanced Scientific Computing Research, Accelerated Research in Quantum Computing under Award Number DE-SC0020316 and DE-SC0025509. EH and XW acknowledge support from NSF CAREER Award CCF-1942837, a Sloan Research Fellowship. GQ acknowledges support from ARO MURI grant W911NF-18-1-0218. This research used resources of the Oak Ridge Leadership Computing Facility, which is a DOE Office of Science User Facility supported under Contract DE-AC05-00OR22725.

\appendix
\section{\label{sec:methodnorm}Resource normalization}
One assumption to note, which the IBM devices satisfy, is that single-qubit gates have the same duration on all qubits. However, this is not the case on all quantum devices.

With the \sherbrooke{} demonstration, since our goal is to compare suppression methods over the same elapsed wall times, we take care to normalize resources between the active suppression methods to allow us to establish comparable trends in the survival probabilities. Due to pulse imperfections, to test the suppression methods on equal footing, we normalize the different DD methods by number of pulses. Because the suppression methods have 4 (XY4), 10 (UR10), 20 (KDD), and 64 (RGA$_{64c}$) pulses per cycle, where possible, we match up the wall time suppression duration of data points for XY4, UR10, and KDD to steps of eighty pulse durations: forty pulses and forty delays. For all DD methods we take data points out to the least common multiple of the DD cycle lengths, 320 physical pulses, making for a suppression duration of 640 pulse durations.

As the metric is so sensitive to a failure on any qubit, for the \kyiv{} demonstration we run with shorter time steps between data points for IDLE and SIM-DD to better observe and more accurately characterize the decay in survival probability. Here, IDLE data is taken for ten points, each separated by eight pulse durations. This puts the suppression durations of IDLE at up to 80 pulse durations. The target for the SIM methods was to reach about 100 physical pulses, so the suppression durations were about 200 pulse durations. Specifically, we ran SIM-XY4 for twelve data points with eight physical pulses per data point, SIM-KDD for five data points with twenty physical pulses per data point, SIM-UR10 for ten data points with ten physical pulses per data point, and RGA$_{64c}$ for two data points with sixty-four pulses per data point.

\section{\label{sec:devices}Devices}
For the demonstrations, we use \kyiv{} and \sherbrooke{} the best-performing devices among IBM's Eagle r3 architecture according to error per logical gate (EPLG) available at compilation time. Similarly, we select \marrakesh{}, a Heron r2 architecture device. Proof of concept tests were conducted on \osaka{}, \kolkata{}, and \cairo{} as well.

\begin{figure*}[t]
    \centering
    \includegraphics[width=0.9\linewidth]{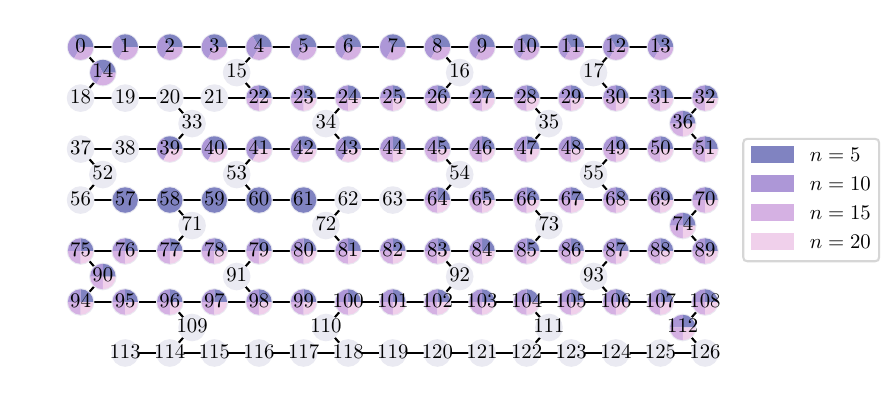}
    \caption{Qubit and coupling map of \sherbrooke{}. Qubits are colored according to whether they were a member of one of the corresponding $n$-qubit embeddings or are gray otherwise.}
    \label{fig:sherbrooke_map}
\end{figure*}

\begin{figure*}[t]
    \centering
    \includegraphics[width=0.9\linewidth]{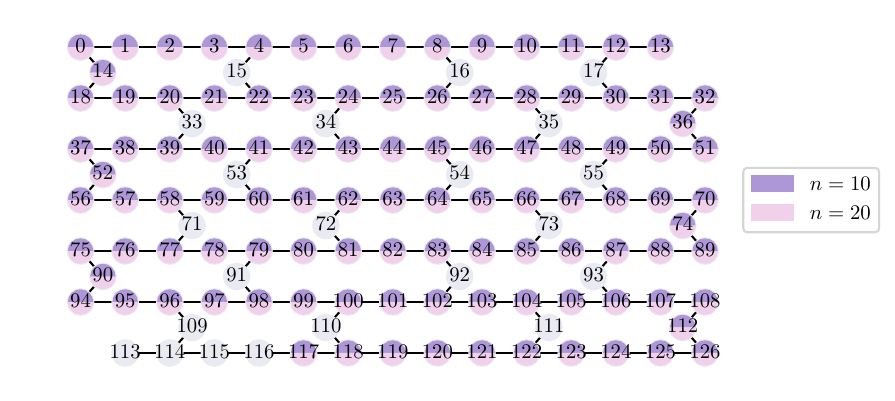}
    \caption{Qubit and coupling map of \kyiv{}. Qubits are colored according to whether they were a member of one of the corresponding $n$-qubit embeddings or are gray otherwise.}
    \label{fig:kyiv_map}
\end{figure*}

Fixing the qubit embedding, initial state, and suppression method, a discontinuity within a trace for increasing protection duration would interfere with the downstream analysis steps. We use the \texttt{qiskit-ibm-runtime} \texttt{Batch} construct to avoid discontinuities due to recalibration within a sequence of circuits so we can fit a function to the results. This construct runs a large set of pre-transpiled jobs back-to-back with minimal interruption. The cloud service intermittently monitors service quality between jobs, an operation outside of our control.

All circuits within a job are run together (up to 300 circuits per job are allowed at time of execution). To take advantage of that, for each qubit group, initial state, and suppression method, we create a job containing circuits for the data points in traces for idle and either all of the crosstalk-robust methods or all of the simultaneous methods.

\subsection{Device Properties}
For the large-scale demonstrations on \kyiv{} and \sherbrooke{}, qubit embeddings were selected by taking continuous paths of qubits (length 5, 10, 15, 20) in the devices' topology graphs. We avoid placing a path over an edge with a 2-qubit gate with reported error rate equal to 100 percent at compile time. This is done in case such an error rate is indicative of a hardware failure that would change the usual accumulation of crosstalk in a way that may unfairly represent the performance of one method or another. 

\begin{figure*}[t]
    \centering
    \includegraphics[width=0.9\linewidth]{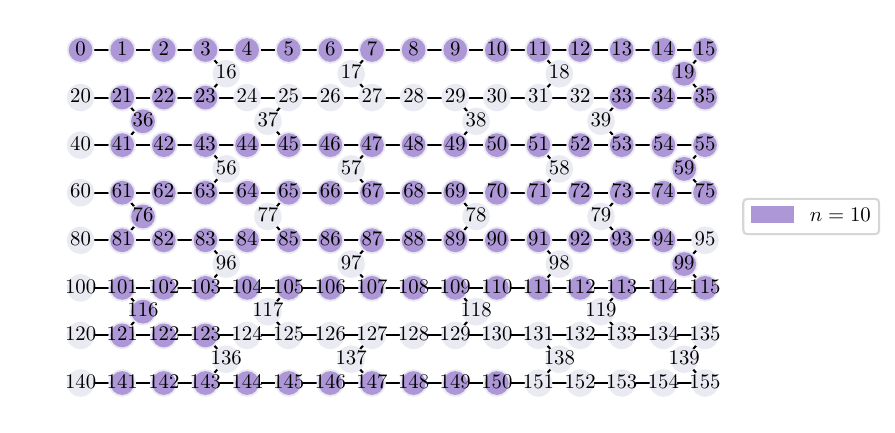}
    \caption{Qubit and coupling map of \marrakesh{}. Qubits are colored according to whether they were a member of one of the corresponding 10-qubit embeddings.}
    \label{fig:marrakesh_map}
\end{figure*}

We select paths both to include a large number of the available qubits in at least one embedding and to reduce overlap between embeddings. The qubits included in the embeddings of each size $n$ are visualized as follows: qubits use on \sherbrooke{} are shown in Fig.~\ref{fig:sherbrooke_map}, \kyiv{} in Fig.~\ref{fig:kyiv_map}, and \marrakesh{} in Fig.~\ref{fig:marrakesh_map}.

\begin{figure*}[t]
    \centering
    \includegraphics[width=0.9\linewidth]{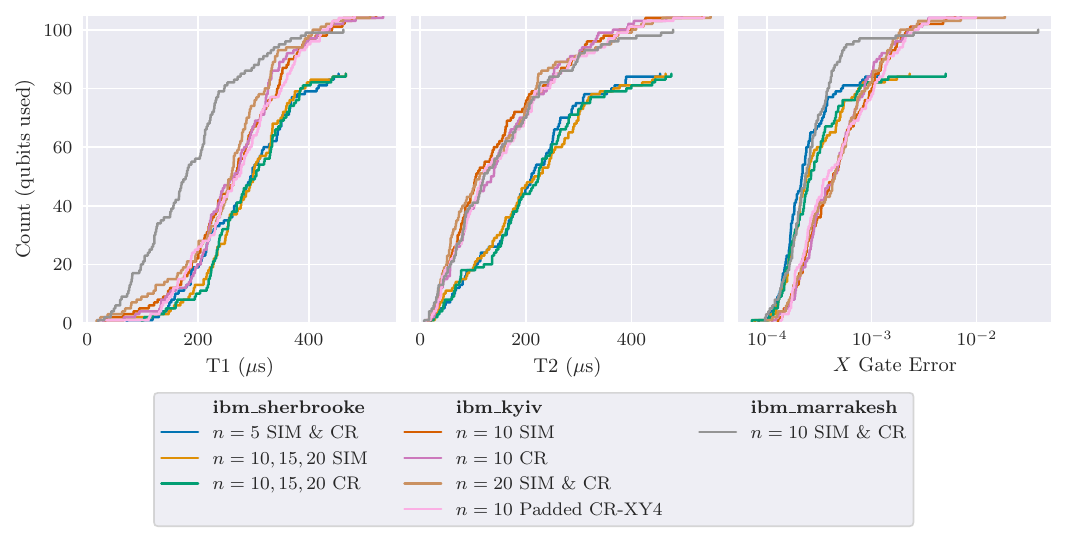}
    \caption{Qubit T1, T2, and $X$ gate error for qubits used in at least one embedding on each device. Curves show the empirical cumulative distribution function of the recorded properties at the time of the execution of the first circuit of a batch job, and are labeled according to the batches of circuits they correspond with.}
    \label{fig:qubit_props}
\end{figure*}

Relevant qubit properties at the time of each batch of experiments are shown in Fig.~\ref{fig:qubit_props} as empirical distribution functions. Since the number of qubits used on each device varies, the $y$-axis is labeled by qubit count rather than proportion or percentage. The demonstrations were completed over 9 different days, so we have included curves showing the properties at the time of experiments for \sherbrooke{}, \kyiv{}, and \marrakesh{}. We ran the full \sherbrooke{} $n=5$ demonstration on 03/27/2024 to get an idea of the device-wide performance. These were followed up by the $n=10,15,20$ SIM-DD and CR-DD demonstrations which were long enough to run on separate days, 04/09/2024, and 04/10/2024, resulting in slightly different qubit properties as a calibration cycle ran between them (device calibration is managed server-side). For the \kyiv{} demonstration, we aimed to run all of the embedding sizes together on 07/04/2024 and 07/05/2024. However, several circuits failed to execute in the $n=20$ batch. The $n=20$ circuits were re-run together on 08/13/2024 and 08/14/2024, but the calibrations on these two dates have the same values for T1, T2, and $X$ gate error, so they are presented as one curve. The demonstration of padding CR-XY4 on \kyiv{} was run on a separate date, 10/22/2024. The \marrakesh{} demonstration was conducted 01/09/2025.

\clearpage

\end{document}